\documentclass[]{aa}

\usepackage{mwe}
\usepackage[utf8]{inputenc}
\usepackage[varg]{txfonts}
\usepackage[T1]{fontenc}
\usepackage{graphicx}
\usepackage{amsmath}
\usepackage{amssymb} 
\usepackage{hyperref}
\usepackage{natbib}
\usepackage{hyperref}
\usepackage{enumerate}
\usepackage{adjustbox}
\usepackage{fancyref}

\usepackage[table]{xcolor}
\definecolor{gray75}{gray}{0.75}
\definecolor{dkgreen}{rgb}{0,0.6,0}
\definecolor{gray}{rgb}{0.5,0.5,0.5}
\definecolor{citationcolor}{rgb}{0,0.5,0}
\definecolor{linkcolor}{rgb}{0.5,0,0}

\hypersetup{colorlinks=true, breaklinks=true, linkcolor=linkcolor, menucolor=linkcolor, urlcolor=linkcolor,citecolor=citationcolor} %for printing add ,hidelinks
\usepackage{longtable}
\usepackage{tabularray}
\usepackage[para]{threeparttable}
\usepackage[para]{threeparttablex}
\addtocounter{footnote}{1}
\usepackage{mathrsfs}
\usepackage{paralist}
\usepackage{url}
\usepackage{siunitx}
\usepackage{setspace}
\usepackage{multicol}
\usepackage[bf]{caption}
\usepackage{booktabs,makecell} %for proper spacing between caption and table and allowance for common formatting of column heads.
\usepackage[]{float}
\usepackage{pdfpages}

\usepackage{graphicx}
\usepackage[varg]{txfonts}
\bibpunct{(}{)}{;}{a}{}{,}

\title{Probing jet dynamics and collimation  in radio galaxies}
\subtitle{Application to NGC\,1052}
\author{Ainara Saiz-P\'{e}rez \inst{\ref{affil:wuerzburg}}
\and Christian M. Fromm \inst{\ref{affil:wuerzburg},\ref{affil:frankfurt},\ref{affil:mpifr}} 
\and Manel Perucho \inst{\ref{affil:valencia},\ref{affil:obsvalencia}}
\and Oliver Porth \inst{\ref{affil:amsterdam}}
\and Matthias Kadler \inst{\ref{affil:wuerzburg}}
\and Yosuke Mizuno \inst{\ref{affil:shanghai},\ref{affil:shanghaicollege},\ref{affil:frankfurt}}
\and Andrew Chael \inst{\ref{affil:princeton}}
\and Karl Mannheim \inst{\ref{affil:wuerzburg}}
}
\institute{%  
  Institut f\"ur Theoretische Physik und Astrophysik, Universit\"at W\"urzburg, Emil-Fischer-Str. 31, D-97074 W\"urzburg, Germany\label{affil:wuerzburg}
  \and%
  Institut f\"ur Theoretische Physik, Goethe Universit\"at, Max-von-Laue-Str. 1, D-60438 Frankfurt, Germany \label{affil:frankfurt}
  \and%
  Max-Planck-Institut f\"ur Radioastronomie, Auf dem H\"ugel 69, D-53121 Bonn, Germany\label{affil:mpifr} 
  \and%
  Departament d'Astronomia i Astrofísica, Universitat de València, C/ Dr. Moliner, 50, 46100 Burjassot, València, Spain \label{affil:valencia}
  \and%
  Observatori Astronòmic, Universitat de València, C/ Catedràtic José Beltrán, 46980 Paterna, València, Spain \label{affil:obsvalencia}
  \and%
  Anton Pannekoek Institute, Science Park 904, 1098 XH, Amsterdam, The Netherlands \label{affil:amsterdam}
  \and%
  Tsung-Dao Lee Institute, Shanghai Jiao Tong University, Shanghai 201210, PR China \label{affil:shanghai}
  \and%
  School of Physics and Astronomy, Shanghai Jiao Tong University, Shanghai 200240, PR China \label{affil:shanghaicollege}
  \and%
  Princeton Gravity Initiative, Princeton University, Princeton NJ 08540, USA \label{affil:princeton}
  }%
  
\begin{document}

 \abstract{
  %\textbf
  %{%
   \textit{Context.} Radio galaxies with visible two-sided jet structures, such as NGC~1052, are sources of particular interest to study the collimation and shock structure of active galactic nuclei jets. High-resolution very-long-baseline interferometry  observations of such sources can resolve and study the jet collimation profile and probe different physical mechanisms.  \newline 
   \textit{Aims.} In this paper, we study the physics of double-sided radio sources at parsec scales, and in particular investigate whether propagating shocks can give rise to the observed asymmetry between jet and counterjet. \newline
   \textit{Methods.} We carry out special relativistic hydrodynamic simulations and perform radiative transfer calculations of an over-pressured perturbed jet. During the radiative transfer calculations we incorporate both thermal and nonthermal emission while taking the finite speed of light into account. To further compare our results to observations, we create more realistic synthetic data including the properties of the observing array as well as the image reconstruction via multifrequency regularized maximum likelihood methods. We finally introduce a semi-automatized method for tracking jet components and extracting jet kinematics. \newline
   \textit{Results.} We show that propagating shocks in an inherently symmetric double-sided jet can lead to partially asymmetric jet collimation profiles due to time delay effects and relativistic beaming. These asymmetries may appear on specific epochs, with one jet evolving near conically and the other one parabolically (width profile evolving with a slope of $\approx1$ and $\approx0.5$, respectively). However, these spurious asymmetries are not significant when observing the source evolve for an extended amount of time.\newline
    \textit{Conclusions.} Purely observational effects are not enough to explain a persisting asymmetry in the jet collimation profile of double-sided jet sources and hint at evidence for asymmetrically launched jets.}

 \maketitle

\section{Introduction}

Radio-loud active galactic nuclei (AGN) contain relativistic jets whose formation and evolution are not yet fully understood (for a review on AGN radio jets, see \citet{2019ARA&A..57..467B}). Both the ergosphere \citep{1977MNRAS.179..433B} and the accretion disk \citep{1982MNRAS.199..883B} of the central black hole of an AGN are thought to play an important role in the formation and launching of relativistic jets. Astrophysical processes are expected to affect the collimation of a jet as it travels through the surrounding medium. This collimation is described by a width profile $w(d) \propto d^k $, where $w$ is the jet width, $d$ is the distance from the central engine, and $k$ is a slope whose value describes a jet that is staying cylindrical ($k=0$), growing parabolically ($k=0.5$), or conically ($k=1$) \citep[see][]{2013ApJ...775...70H,2020MNRAS.495.3576K,2021A&A...647A..67B,2022A&A...664A.166R}.

Ever since the computational power was available, numerical special relativistic hydrodynamic (SRHD) simulations have been used as a tool to study the parsec and sub-parsec scales of relativistic jets. Initial studies focused on the collimation  and expected radio emission of steady-state jets \citep{1996ASPC..100..173K}, as well as on the effect of the ambient medium on their standing shock structure \citep{1995ApJ...449L..19G}. Later studies found that perturbing the jet inlet gave rise to travelling shocks, resulting in bright components, observable at apparent superluminal speeds at radio frequencies \citep{1997ApJ...482L..33G,1997MNRAS.288..833K}. These travelling shocks affect the dynamics of the jet, particularly in their interactions with the standing shocks. These works found that when a travelling shock passes through a standing shock, this standing shock is dragged downstream before going back to its equilibrium position. Furthermore, the initially bright and compact shocks become extended and much less bright upon crossing a standing shock, and their width affects the collimation profile of the jet downstream. For all of this to be studied, they determined that the aberration seen by an observer due to the finite speed of light was absolutely necessary. This was confirmed by later studies, which also discovered that sufficiently strong perturbations will cause trailing shocks \citep{Agudo2001}. These shocks travel behind a main component with a slower velocity, and may not be observable at low radio frequencies due to the large blurring beams \citep{2009ApJ...696.1142M}. \citet{2016A&A...588A.101F} studies in detail the interaction between  travelling shock waves and recollimation shocks using special relativistic hydrodynamic simulations and radiative transfer calculations. The study focuses on the impact of travelling perturbations on the spectral evolution of AGN jets at small inclination angles, so-called blazars, at GHz frequencies, as they induce observational signatures that could be detected in very long baseline interferometry (VLBI) observations. 

Due to its large viewing angle of $\theta >$ 80$^{\circ}$ \citep{2022A&A...658A.119B}, meaning that the jet is closely aligned with the plane of the sky, as well as the fact that both its jet and counterjet are visible, the low-luminosity AGN in the galaxy NGC~1052 is a source of particular interest to study jet properties. The source is located at a redshift of $z=0.005037$ \citep{2003ApJ...583..712J} ($D\approx19.5$~Mpc) and is classified as a  low-ionization nuclear emission line region (LINER) galaxy due to its nuclear properties \citep{1939PASP...51..282M,1978MNRAS.183..549F,1997ApJS..112..315H}. It hosts a supermassive black hole with a mass of $\approx 10^{8.2}$~$\mathrm{M_\odot}$ \citep{2002ApJ...579..530W}, though more recent works estimate a greater value of $\approx10^{9.3}$~$\mathrm{M_\odot}$ \citep{2020ApJ...895...73K}.

Observations of the source done at frequencies around 1.5~GHz revealed a double-sided jet structure spanning nearly 3~kpc, with a central, flat-spectrum compact core containing around 85\% of the emission flux \citep{1984ApJ...284..531W,2004A&A...420..467K}. 
Very long baseline interferometry studies estimate a position angle\footnote{we define the angle clockwise as measured from the x-axis} of the source in the plane of the sky between 10$^\circ$ and 25$^\circ$ \citep{2004A&A...420..467K,2016A&A...593A..47B,2020AJ....159...14N}. In this work, we set this position angle to 15$^\circ$. Images by the Very Long Baseline Array (VLBA) and the Global mm-VLBI Array (GMVA) also show a prominent emission gap between jet and counterjet at cm wavelengths, decreasing in size with increasing frequency and vanishing for $\nu \geq 43$~GHz \citep{2003A&A...401..113V,2003PASA...20..134K,2004A&A...426..481K,2016A&A...593A..47B}. Modeling studies show this sort of gap could be caused by a torus-like structure presenting free-free absorption \citep{2018A&A...609A..80F}. 

Sub-parsec scale studies of the source done at 15~GHz between 1995 and 2001 by the VLBA estimated the outward motions in the jets to have a roughly equal apparent velocity of $0.26\pm0.04\,c$ \citep{2003A&A...401..113V}. These velocities are in agreement with a detailed 15~GHz analysis of MOJAVE data between 1995 and 2012 that revealed an average velocity of the components of $0.230\pm0.011\,c$ \citep{2013PhDT.......479B}. \citet{2019A&A...623A..27B} calculate faster mean apparent velocities through VLBA observations done at 43~GHz between 2004 and 2009, $0.529\pm0.038\,c$ for the eastern jet and $0.343\pm0.037\,c$ for the western counterjet. An asymmetry was also found in the brightness of the jet and counterjet, starting in 2007, while further studies \citep[see][]{2022A&A...658A.119B} find asymmetries between the collimation profiles.

Such apparent asymmetry could arise from a symmetric source due to the effects of an obscuring torus, as well as due to relativistic beaming of the misaligned jet-counterjet system. \citet{2019A&A...623A..27B} could not find consistent parameters for the intrinsic velocities and viewing angle of all observations, pointing to an intrinsic asymmetry between the jet and the counterjet. While the majority of models for relativistic jet launching result in symmetric outflows, several works have investigated intrinsically asymmetric launching. One such proposed setup consists of the torus around a rotating black hole being seeded with a multiloop magnetic field configuration \citep[see][]{2015MNRAS.446L..61P,2020MNRAS.494.4203M}. \citet{2020MNRAS.495.1549N}, \citet{2021MNRAS.508.1241C}, and \citet{Jiang2023} run general relativistic magnetohydrodynamic simulations of such setup, and find that an alternating polarity in the magnetic loops results in intermittent jets. This effect may be severe enough that the launching of the counterjet is inhibited. \citet{2013ApJ...774...12F} studied several alternative methods for asymmetric jet launching. They found that a sophisticated prescription of magnetic diffusivity in the torus, accounting for position and local sound speed, resulted in the longest-lasting asymmetry between the jet and the counterjet.

The aim of this work is to study the possibility and extent of inducing perceived asymmetries on a symmetric jet-counterjet system at parsec and sub-parsec scales. We carried out a 2D-axisymmetric SRHD simulation of a mirrored jet in an ambient medium with a decreasing pressure gradient. Once the jets reached a steady state, we injected perturbations into the jet nozzle and had them evolve alongside the jet spine. We then performed radiative transfer calculations on the hydrodynamic results, incorporating both thermal absorption by a torus-like structure and nonthermal emission from the jets. To understand if an asymmetry between both jets could be caused by observational effects, we computed the effect that a finite speed of light has on the emission maps. In addition, to account for the effects of limited resolution and observing time on observational data, we generated synthetic images that recreate real observing conditions. We build on the work carried out by \citet{2019A&A...629A...4F}, incorporating into our method the injection of perturbations into the jet nozzle as well as the multifrequency functionalities of the imaging algorithm \texttt{eht-im} \citep{2023ApJ...945...40C}.

To scale our results to NGC\,1052, we used the redshift-independent distance of $D=19.23\pm0.14$~Mpc, which translates to a linear scale of $0.093$~pc/mas \citep{2020PhDT........17B}.  This paper is organized as follows: in Sect.~\ref{sims} we introduce our numerical setup, breaking it down into the jet-torus setup and SRHD simulation in \ref{rhd} and the radiative transfer in \ref{radtrans}. In Sect.~\ref{synth} we explain our method for generating synthetic images that mimic observational conditions. In Sect.~\ref{analysis} we analyze the final images, and we discuss the results in Sect.~\ref{discussion}.

\section{Numerical simulations} \label{sims}
\subsection{Numerical simulations} \label{rhd}
We use the high-resolution shock-capturing code \texttt{Ratpenat} 
\citep[for more details, see][]{Perucho2010} to compute 2D special relativistic hydrodynamical jet simulations in cylindrical coordinates $(r,z)$, on which we test our synthetic imaging and analysis pipeline. The setup of these simulations can be found in \cite{2018A&A...609A..80F, 2019A&A...629A...4F}. 
During the simulation, we set the speed of light to $c=1$ and the units of length, velocity, density, and pressure as $R_j$, $c$, $\rho_{a,0}$ and $\rho_{a,0}c^2$, respectively. $R_j$ is the jet radius and $\rho_{a,0}$ is the initial density of the ambient medium.
For the sake of completeness, we repeat the basic setup below. Our jet is embedded in an isothermal ambient medium with decreasing pressure following:
\begin{equation}
p_a(z)=\frac{p_{b,0}}{d_k}\left[1+\left(\frac{z}{z_c}\right)^n\right]^{-{m/n}}.
\end{equation}
In the equation above, $p_a$ is the ambient pressure, $p_{b,0}$ is the jet pressure and $d_k$ is the over-pressure of the jet with respect to the ambient at injection. Furthermore, $z_c$ is the core distance and the exponents $n$ and $m$ control the decay of the ambient pressure with distance $z$. We use  $z_c=10\,R_j$, $n=1.5$, and $m=2$. The density of the ambient medium at $z=0$ is $\rho_{a,0}=1.0$ and it decreases with distance. The jet is characterised by its pressure, $p_{b,0}=0.003$, density, $\rho_{b,0}=0.04$, velocity, $v_{b,0}=0.5\,c$ at injection indicated by the subscript `0' and the overpressure $d_k=2.5$. An overpressured jet in a decreasing-pressure ambient medium results in a jet with a series of recollimation shocks, which are identified as the bright knots observed by VLBI techniques \citep[e.g.][]{1997ApJ...482L..33G,2018A&A...609A..80F}. Finally, we assume an ideal equation of state using an adiabatic index of $\hat{\gamma}=13/9$, the expected one for relativistic electrons and cold protons at the same temperature. \\
To mimic an obscuring torus we include a flared disk characterized by its inner and outer radii, $R_
\mathrm{in,out}$, angular thickness, $\Theta$, density $\rho(\rho_{R_{\mathrm{in}}},r,\theta)$ and temperature profile $T(T_{R_{\mathrm{in}}},r,\theta)$ \citep[for details see][]{2018A&A...609A..80F}. In Table \ref{table:torus} we list all parameters used for the simulation.

\begin{table}[h!]
\centering
\caption{Parameters used for the simulation}
\label{table:torus}
\begin{tabular}{c c c c }
\hline\hline
\\[-0.9em]
$\rho_{\rm b,0}$ $[\rho_{a}]$ & $p_{\rm b,0}$ $[\rho_{a}c^2]$ & $v_b$ [c]  &$\hat{\gamma}$ [1] \\
\\[-0.9em]
\hline
\\[-0.8em]
0.04 & 0.003 & 0.5 & 13/9 \\
\hline \hline
\\[-0.9em]
$z_{\rm c}$ $[R_{\rm j}]$ & $n$ $[1]$ & $m$ [1]  & $d_k$ [1] \\
\\[-0.9em]
\hline
\\[-0.8em]
10 & 1.5 & 2 & 2.5 \\
\hline \hline
\\[-0.9em]
$R_{\rm in}$ $[R_j]$ & $R_{\rm out}$ $[R_j]$ & $\Theta$ [deg]  & $\rho_{R_{\mathrm{in}}}$ $[\rho_a]$  \\
\\[-0.9em]
\hline
\\[-0.8em]
4 & 30 & 50 & 70 \\
\hline \hline
\\[-0.9em]
$T_{R_{\mathrm{in}}}$ [K]  & k$^a$ [1] & l$^b$ [1] &\\ 
\\[-0.9em]
\hline
\\[-0.8em]
1500 & 1 & 2 \\ \hline
\multicolumn{4}{l}{$^a$ controls the radial profile $\propto r^{-k}$}\\
\multicolumn{4}{l}{$^b$ controls the polar profile $\propto e^{-l|\cos\theta|}$}\\
\end{tabular}
\end{table}

\begin{figure}[h!]
\includegraphics[width=9cm]{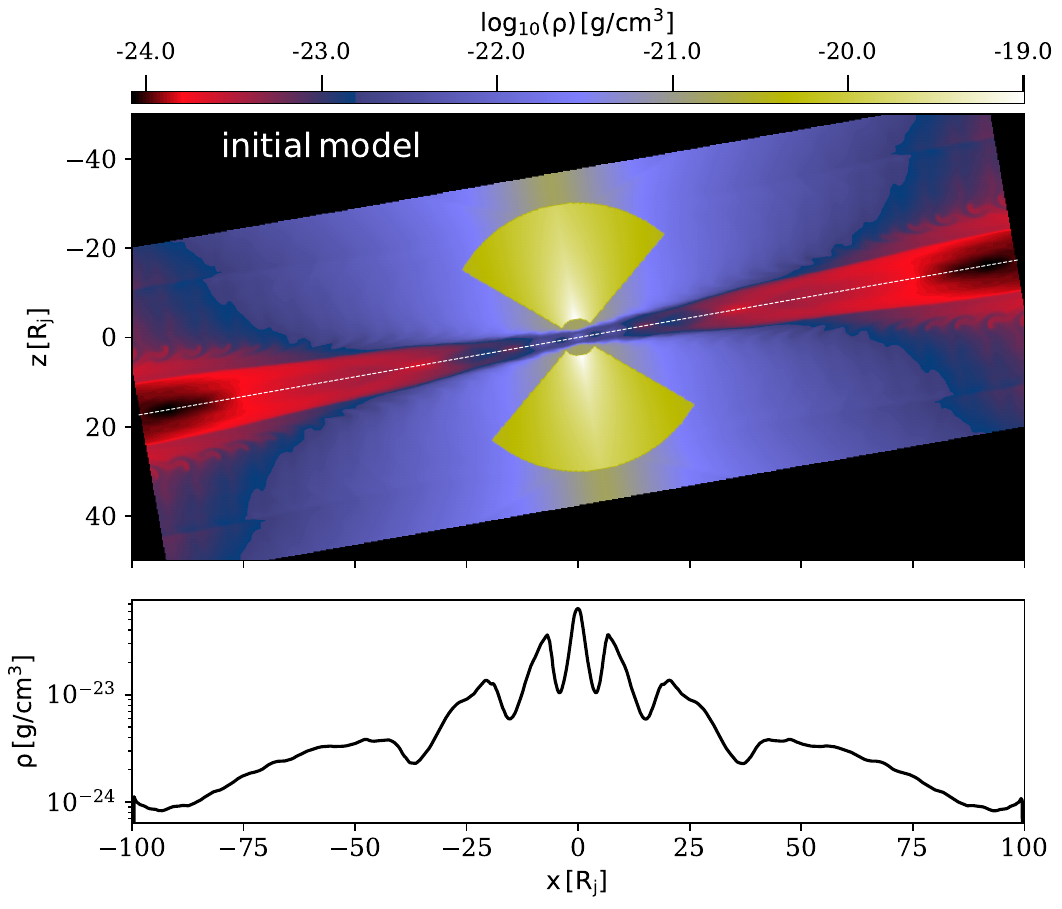}
\caption{Distribution of the density of the initial steady-state jet model in a slice through the mid plane $(y=0)$ of the jet. The jet is seen under a viewing angle of $\vartheta=80^
\circ$. The top panel shows the 2D density distribution and the bottom panel a density evolution along the jet axis (dashed white line).}
\label{fig:inithydro}
\end{figure}

The numerical grid covers $80\times100\,R_j$, using 4 cells per jet radius at the injection nozzle. The initial overpressure leads to a pinching jet and triggers the formation of recollimation shocks \citep[see, e.g.,][]{2016A&A...588A.101F}. The steady state jet structure is obtained after roughly five longitudinal light crossing times and is presented in Fig. \ref{fig:inithydro}.  The bottom panel of Fig. \ref{fig:inithydro} displays the density evolution along the jet axis (dashed white line in the top panel). As can be seen from the local density maxima there are several recollimation shocks located at $x=\pm7,\pm21,\textrm{ and }\pm42\,R_{\rm j}$. The distance between the recollimation shocks increases while their density jumps decrease with distance. This behaviour is expected from relativistic jets evolving in a decreasing pressure atmosphere \citep[see, e.g.,][]{2018A&A...609A..80F}. 

In order to study the jet dynamics, including the propagation of shocks, we inject periodic density and pressure variations into the jet nozzle with a maximum amplitude of four times the injection values. These perturbations lead to the formation of travelling shock waves. We generate snapshots from the SRHD simulations every $t=0.25R_j/c$ time steps\footnote{the time step is determined by the cell crossing time}  for later use during the radiative transfer calculations.  

\subsection{Radiative transfer} \label{radtrans}
We use the produced SRHD snapshots to compute the jets' radiative output, assuming nonthermal synchrotron emission from the jet (including self-absorption) and free-free absorption in the torus. In addition, we take the finite speed of light into account, i.e. slow-light approach. 

As we are not evolving a nonthermal particle distribution within our numerical models, we instead relate them to the density and energy of the thermal particles. We assume a power-law distribution of nonthermal particles:

\begin{equation}
  n(\gamma_e)=\begin{cases}
    n_0\left(\frac{\gamma_e}{\gamma_{e,\rm min}}\right)^{-s} & \text{if $\gamma_{\rm e,min}\leq\gamma_{e}\leq \gamma_{\rm e,max}$}\\
    0  & \text{otherwise},
  \end{cases}
\end{equation}
where $\gamma_e$ is the electron Lorentz factor, $n_0$ is the normalization coefficient and $s$ is the spectral slope. The normalization coefficient and the boundaries of the particle distribution, $\gamma_{\rm e,min/max}$, can be obtained assuming that the nonthermal particles are a fraction $\zeta_e$ of the thermal particles, containing a fraction $\epsilon_e$ of their energy. Since we are not evolving the magnetic field, we assume a fraction $\epsilon_b$ of the equipartition magnetic field during the emission calculations \citep[for details, see][]{2019A&A...629A...4F}. Given that the SRHD simulations are scale-free, we scale them to cgs units by providing a jet radius, $R_j$, in cm and an ambient density, $\rho_a$, in g/cm$^3$. Moreover, we assume a viewing angle with respect to the line of sight of $\vartheta=80^\circ$ and a redshift of $z=0.005$. In Table \ref{table:emiss} we list the parameters used to scale our model, chosen to match the total flux of the reference source NGC~1052 at the central frequency of 22~GHz \citep{2004A&A...426..481K}. Given NGC~1052's linear scale of 0.093 pc/mas, the jet scale is $\approx 2\cdot 10^{16} $ cm per jet radius. This is compatible with the length scale in Table~\ref{table:emiss}, as well as the estimated upper limit for the jet core size at 86~GHz \citep{2016A&A...593A..47B}.

\begin{figure}[h!]
\includegraphics[width=9cm]{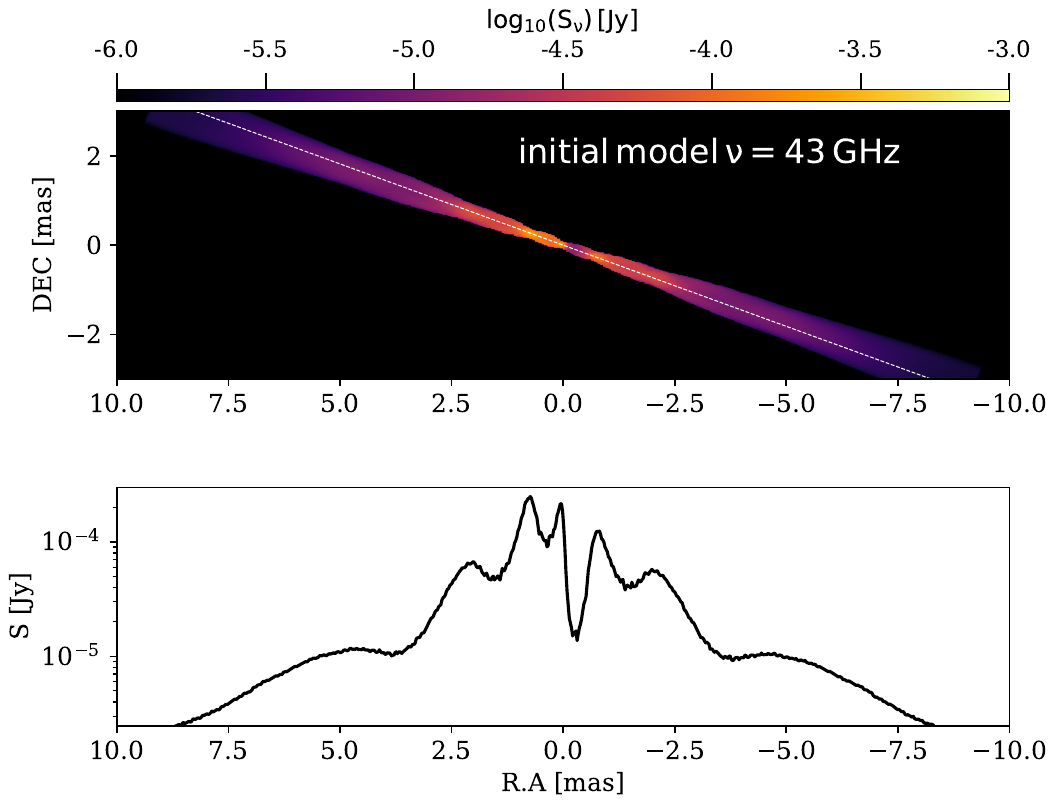}
\caption{Emission from the initial steady-state jet at 43\,GHz. The top panel displays the jet structure and the bottom panel the evolution of the flux density along the jet axis (white dashed line)}
\label{fig:initemission}
\end{figure}

\begin{table}[h!]
\centering
\caption{Parameters used for the emission calculations}
\label{table:emiss}
\begin{tabular}{c c c c c c c}
\hline\hline
$R_{j}$& $\rho_{a}$ & $\zeta_e$   & $\epsilon_e$  &  $\epsilon_b$  & $s$\\ \hline
[10$^{16}$ cm] & [10$^{-21}$ g/cm$^3$] & [1] & [1] & [1] & [1] \\ \hline
3 & 1.6 & 0.6 & 0.05 & 0.1  & 3.5\\ \hline
\end{tabular}
\end{table}
The initial 2D axisymmetric simulations are interpolated to a 3D cartesian grid via a Delaunay triangulation, assuming a viewing angle $\vartheta$. During the Delaunay triangulation, we store the vertices and their weights. Once they are computed and kept in memory, additional triangulations represent an interpolation of data on the 3D cartesian grid using the precomputed vertices and weights. This procedure significantly speeds up the radiative transfer calculations \citep[for details on the numerical grid and conversion studies, see][]{2018A&A...609A..80F}.
The emission structure of the initial jet at 43\,GHz is shown in Fig.~\ref{fig:initemission}. Notice that the radiative transfer leads to a projection of the source structure onto the x-y plane as seen in Fig.~\ref{fig:inithydro}. This image is then rotated by $\phi=15^\circ$ in the plane of the sky to match the orientation of NGC\,1052. At 43\,GHz the obscuring torus is almost entirely optically thin, therefore the jet nozzle becomes visible in the emission map. However, due to the slight misalignment of the jet with respect to the observer ($\vartheta=80^\circ$) the absorption along the line of sight is greater for the right jet, which manifests as a dimmer innermost jet structure. A more detailed view of the emission structure is presented in the bottom panel of Fig.~\ref{fig:initemission} which displays the flux density along the jet axis (dashed white line in the top panel). The aforementioned recollimation shocks are clearly visible and can be found at a right ascension of $\mathrm{R.A.}=\pm0.8, \pm2.1,\mathrm{ and } \pm4.5 \mathrm{~mas}$. Notice that the last recollimation shock resembles a plateau rather than a local flux density maxima, as would be typical of a strong recollimation shock.

We initialise the slow-light calculations by using the Delaunay triangulation of the last SRHD snapshot and search for the jet location closest to the observer. 

\begin{figure}[h!]
\includegraphics[width=9cm]{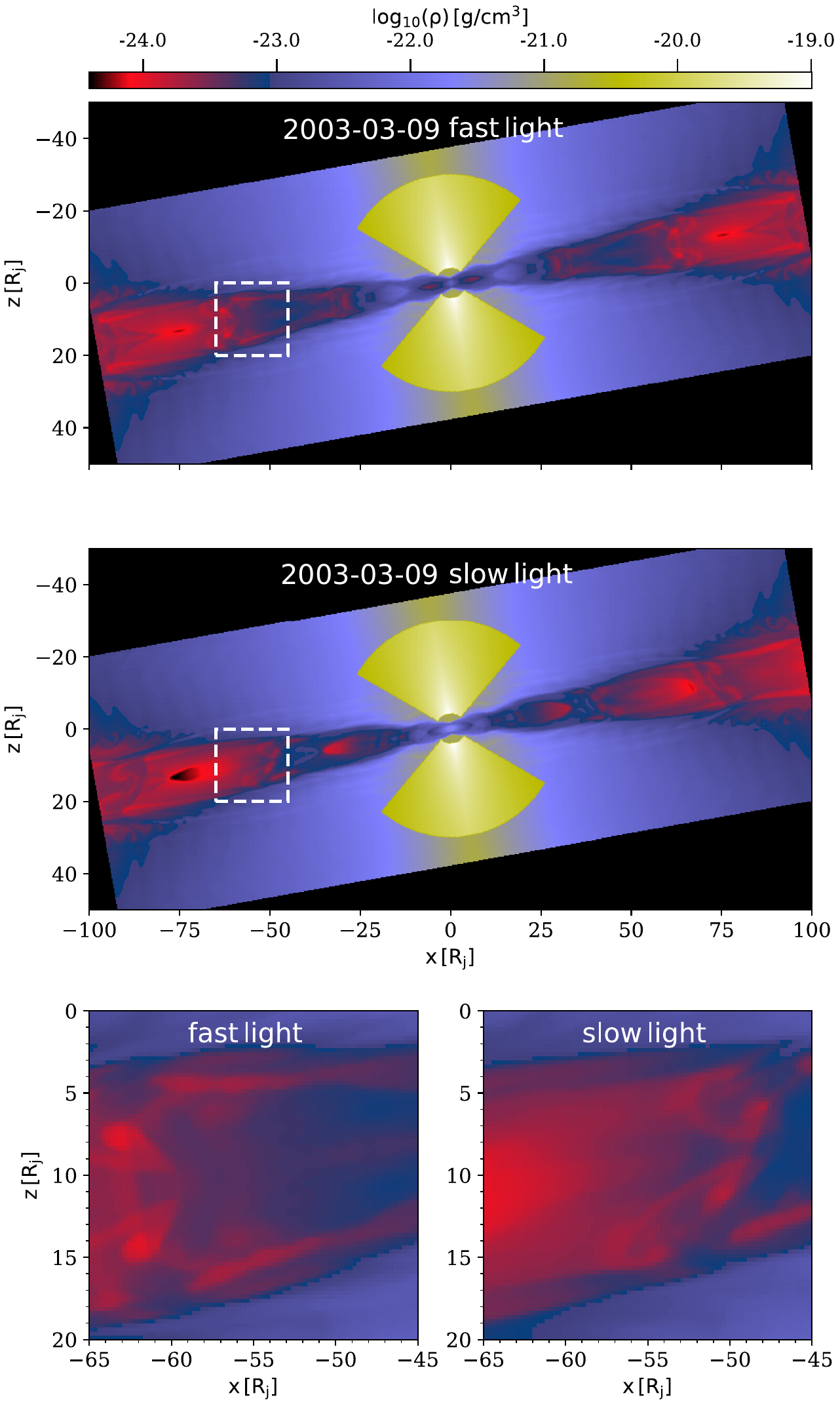}
\caption{Distribution of the density in a slice through the mid plane $(y=0)$ of the jet. The jet is seen under a viewing angle of $\vartheta=80^
\circ$ and the radiative transfer is performed in a positive z-direction. The top panel shows the density without time delays and the middle one including time delays, where the left jet is beamed toward and the right jet away from the observer. The white boxes correspond to the zoom windows shown in the two bottom panels. Notice the rotation of the travelling shock in slow-light case (bottom right panel).}
\label{fig:fastslow}
\end{figure}

From this position on, we integrate backward in time and load the corresponding SRHD snapshots including the plasma parameters. Since we keep the vertices and weights of the Delaunay triangulation in memory, the slow-light calculations can be performed in a fast and efficient way.
In Fig.~\ref{fig:fastslow} we show the density distribution in the mid plane (x-z plane for $y=0$) as seen by the photons for the slow-light (bottom) and for the fast-light (top) approaches. Notice the rotation and broadening of the shocks caused by the light travel time delay between the near side (front) and the far side (back) of the perturbation. These time delays are also responsible for the asymmetry between the left and right jet, which are respectively beamed toward and away from the observer \citep[see also][]{Aloy2003}.

\begin{figure}[h!]
\includegraphics[width=9cm]{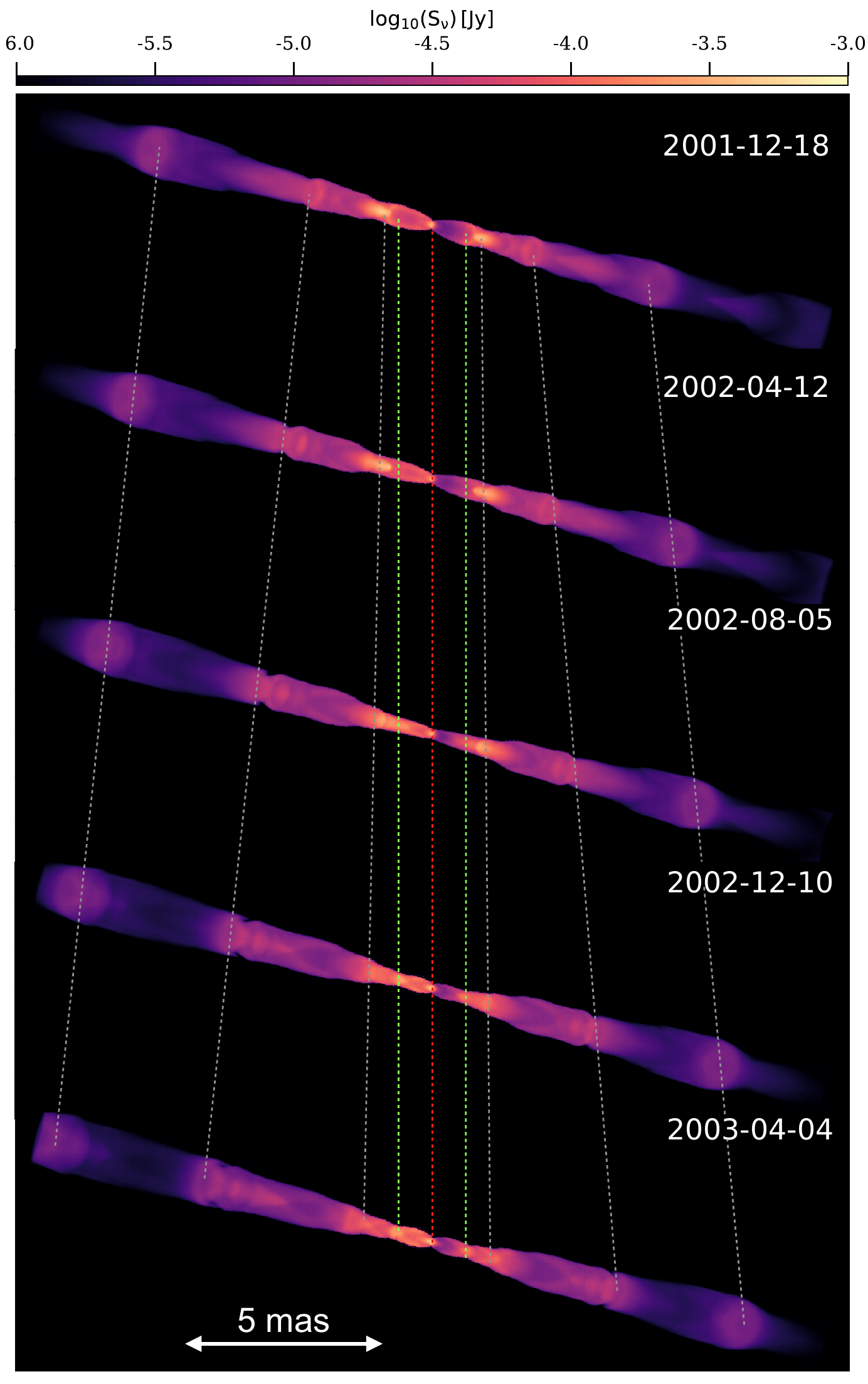}
\caption{Evolution of the jet structure at 43\,GHz for different time frames $\Delta t\sim115\,\mathrm{days}$. The red line indicates the location of the jet origin and two green lines mark the position of the first recollimation shock. The gray lines track the evolution of propagating shock waves.}
\label{fig:timeevo}
\end{figure}

In Fig.~\ref{fig:timeevo} we show the emission structure obtained from the simulations at 43\,GHz for several snapshots separated by $\Delta t=10R_j/c\approx115\,\mathrm{days}$. Notice that the torus turns optically thin at $\nu\sim 30\,
\mathrm{GHz}$ and thus we can clearly see the jet injection point (dotted red line) together with the first recollimation shocks (dotted green lines). Several travelling shock waves are visible and traceable (dotted gray lines). A deeper look into the structures of the travelling shocks reveals trailing shocks propagating in the wake of the main (forward) shock \cite[see, e.g.,][]{Agudo2001,2022A&A...661A..54F}.\\

\begin{figure}[h]
\includegraphics[width=9cm]{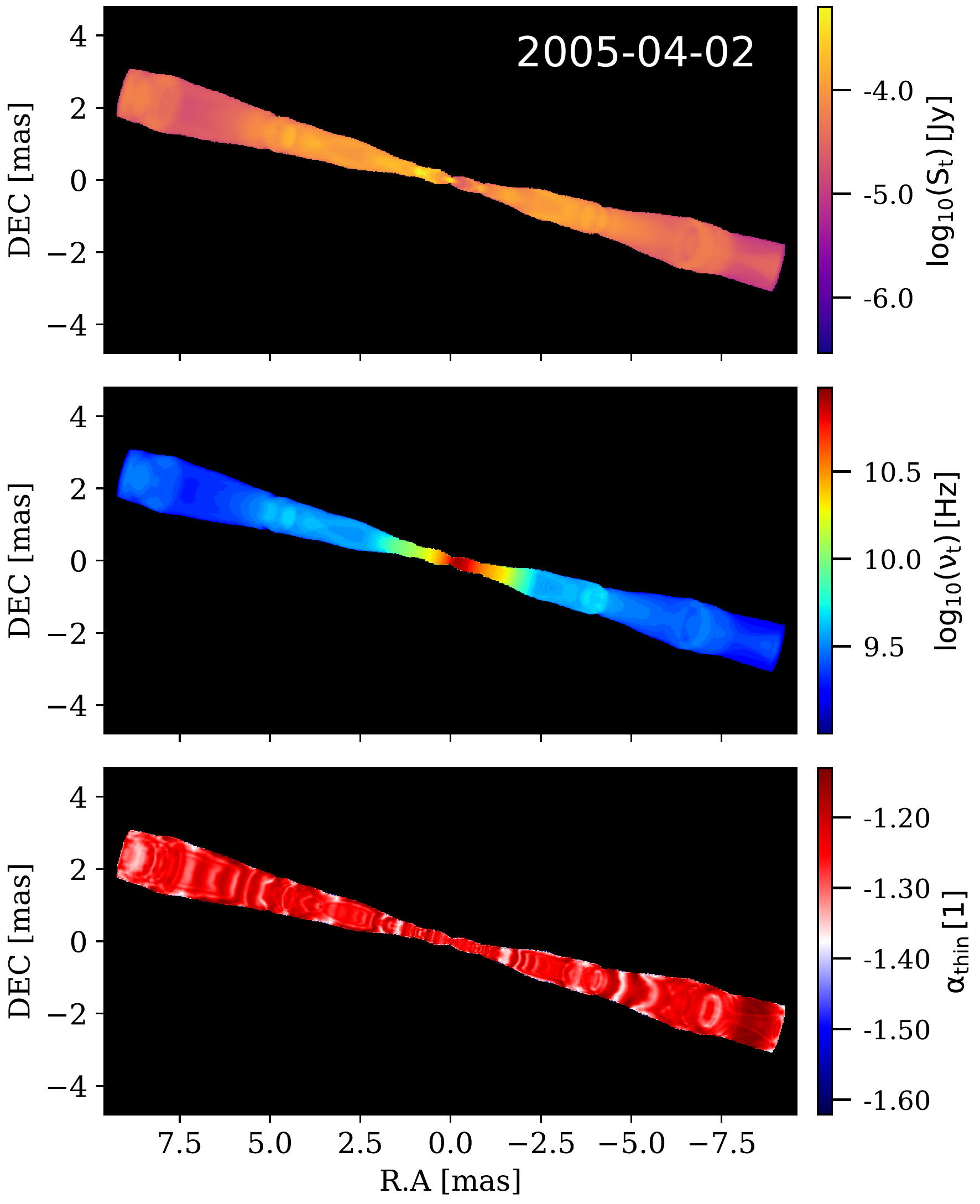}
\caption{Spatial evolution of the spectral turnover position for a selected time frame. The panels show from top to bottom the turnover flux density $S_t$, the turnover frequency $\nu_t$ and the optically thin spectral index $\alpha_{\rm thin}$.}
\label{fig:turnover}
\end{figure}

Besides the temporal and spatial variations of the jet structure for a fixed frequency, our larger frequency coverage allows us to trace the spectral properties, e.g. the turnover flux density $S_t$, the turnover frequency $\nu_t$ and the optically thin spectral index $\alpha_{\rm thin}$. Given the variation of the spectral index together with synchrotron self-absorption (SSA) in the jet and free-free absorption in the torus, we compute the optically thin spectral index from our last frequency bin, i.e. between 900\,GHz and 1\,THz. At this frequency range SSA and free-free absorption do not affect the spectrum. In Fig. \ref{fig:turnover} we present from top to bottom the turnover flux density $S_t$, the turnover frequency $\nu_t$ and the optically thin spectral index $\alpha_{\rm thin}$. The variation of the turnover position can clearly be seen from the plots. Due to the obscuring torus and high density and pressure at the injection region the largest values for the turnover flux density and turnover frequency are located within $\pm 1$\,mas from the origin. 

In addition, the local maxima in the turnover position nicely trace the moving shocks. Since we do not include radiative cooling in our emission modeling, the increase of the turnover values are due to shock compression of the underlying flow. The median optically thin spectral index (bottom panel) is $\langle \alpha_{\rm thin}\rangle =-1.25\pm0.1$ which corresponds to the expected values for the used spectral slope of $s=3.5$. During the radiative transfer we numerically solve the integrals over the Bessel functions and do not use approximations. This leads to a deviation from the expected spectral index in the high-frequency emission which explains the variation of the spectral index.

\begin{figure}[h]
\includegraphics[width=9cm]{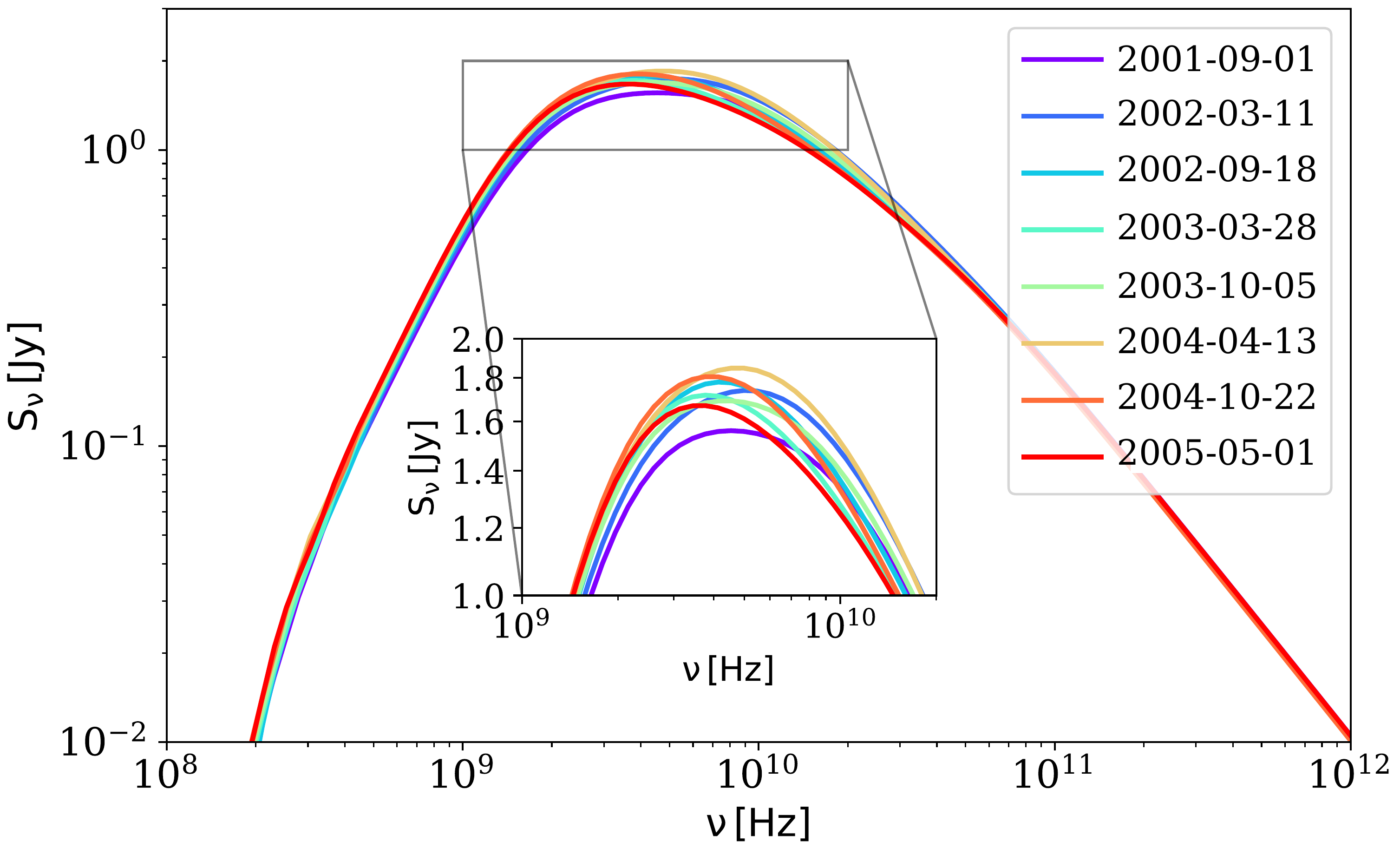}
\caption{Temporal evolution of the broadband spectrum for our model. The different colors indicate selected time steps. }
\label{fig:spectrum}
\end{figure}

The unresolved total spectrum between $10^8\,\mathrm{Hz}<\nu<10^{12}\,\mathrm{Hz}$ and its temporal variation over several time frames is shown in Fig.~\ref{fig:spectrum}. The inlet provides a zoom into the turnover location showing the variation of the spectrum during the course of the simulation, which is connected to the propagation of perturbations and their interactions with the recollimation shocks.

\section{Synthetic data generation} \label{synth}
\subsection{Setup}

To better compare the output of our simulations with observations, we aim to obtain a data set that incorporates hurdles present in VLBI results. These include a sparse sampling of the $u-v$ plane as a result of  a limited number of telescopes, the presence of noise, and the effects of image reconstruction. These effects are not fully considered when blurring the emission maps from a simulation with an observing beam, as is typically done to imitate real observations. To put our data set through an image reconstruction process, we used the synthetic observation tools of the imaging code \texttt{eht-im} \citep{2016ApJ...829...11C,2018ApJ...857...23C}.

We first decided on an observing array as well as an observation schedule to generate the projected baselines of a synthetic observation for 190 consecutive outputs. In this work, we use the characteristics of the VLBA at five different observing frequencies, as well as certain GMVA stations for 86~GHz. The VLBA consists of 10 identical 25~m antennas spread over the US, with a longest baseline of around 8600~km. Multifrequency observations can be carried out in different bands ranging from 0.3~GHz to 90~GHz, with 8 out of the 10 telescopes being able to observe in the band centered at 86 GHz. The telescope sensitivities, given in system equivalent flux density (SEFD)\footnote{data taken from \url{https://science.nrao.edu/facilities/vlba/docs/manuals/oss/bands-perf}}, of the VLBA corresponding to the observing frequencies used in this work are shown in Table~\ref{table:SEFDVLBA}. We also included GMVA\footnote{data taken from 
 \url{https://www3.mpifr-bonn.mpg.de/div/vlbi/globalmm/sensi.html}} stations at 86~GHz, as given in Table~\ref{table:SEFDGMVA}. For this work, we used the VLBA bandwidth of $\Delta \nu=128$\,MHz.

\begin{table}[]
\centering
\caption{SEFD for VLBA telescopes at different frequencies}
\label{table:SEFDVLBA}
\begin{tabular}{c c c c c c}
\hline\hline
frequency {[}GHz{]} & 8   & 15  & 22  & 43   & 86                                                    \\ \hline
SEFD {[}Jy{]}       & 327 & 543 & 640 & 1181 & 2500 \\ \hline
\end{tabular}
\end{table}

\begin{table}[]
\centering
\caption{SEFD at 86~GHz for the GMVA telescopes}
\label{table:SEFDGMVA}
\begin{tabular}{c c c c c c}
\hline\hline
Station & EF   & PdB  & PV  & ON   & MH                                                    \\ \hline
SEFD {[}Jy{]}       & 1000 & 244 & 654 & 3878 & 20000 \\ \hline
\end{tabular}
\end{table}

\begin{figure}[h]
\centering
\includegraphics[width=9cm]{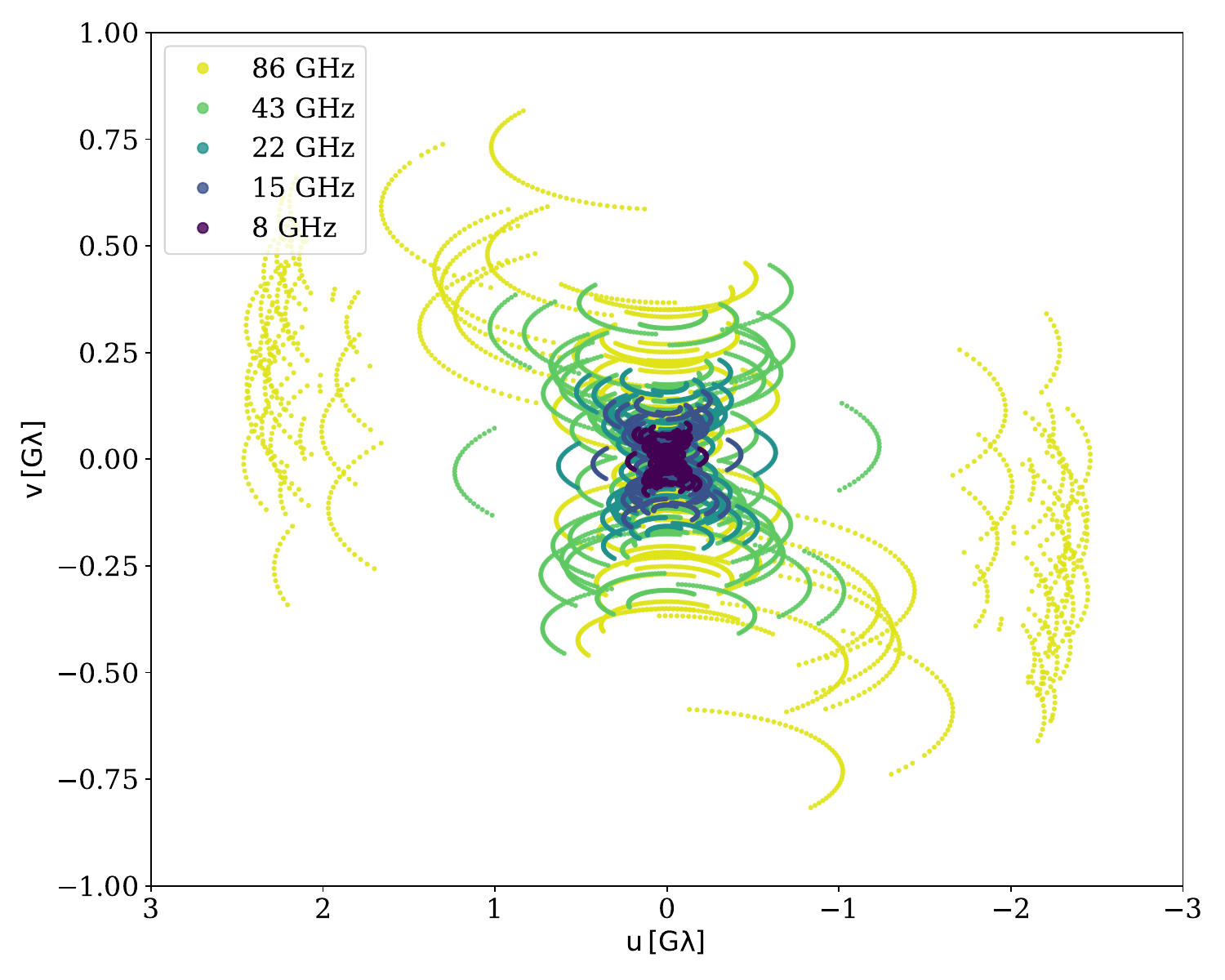}
\caption{$u-v$ coverage of a synthetic observation on 2003 October 9 at different frequencies. Note the different baselines for 86~GHz, due to the additional GMVA stations as well as the VLBA stations not operational at 86~GHz. Notice the different x- and y-scales.}
\label{fig:uv_coverage}
\end{figure}

With the information given in Tables~\ref{table:SEFDVLBA} and ~\ref{table:SEFDGMVA}, as well as as the coordinates of the stations, we computed the projected baselines for a synthetic observation with a duration of 15 hours (typical observing time for NGC\,1052), a scan integration time of 12~s and a cadence of 10 minutes between scans. As per \citet{2019A&A...629A...4F}, we furthermore considered that the telescopes could observe at elevations between 10° and 85°. The time elapsed between each consecutive output was $\Delta t=R_j/4c\approx3\,\mathrm{days}$. We started on 2003 October 9 for our first set of data and calculated the starting hour of the observation for each time step. For this purpose, we chose the VLBA station St. Croix as a reference antenna. From these parameters, we generated projected baselines, with Fig.~\ref{fig:uv_coverage} showing the resulting $u-v$ coverage of one such synthetic observation at each of our working frequencies. We then generated synthetic visibilities by Fourier transforming and sampling the flux density distribution, $I(x,y)$, obtained in Sect.~\ref{radtrans} with the aforementioned baselines. This is done through the equation  

\begin{equation}\label{eq:visibilities}
V_{ij} =\int\int I(x,y)e^{-2\pi \textrm{i}(ux+vy)}\textrm{d}x\textrm{d}y.
\end{equation}

\begin{figure}[h]
\centering
\includegraphics[width=9cm]{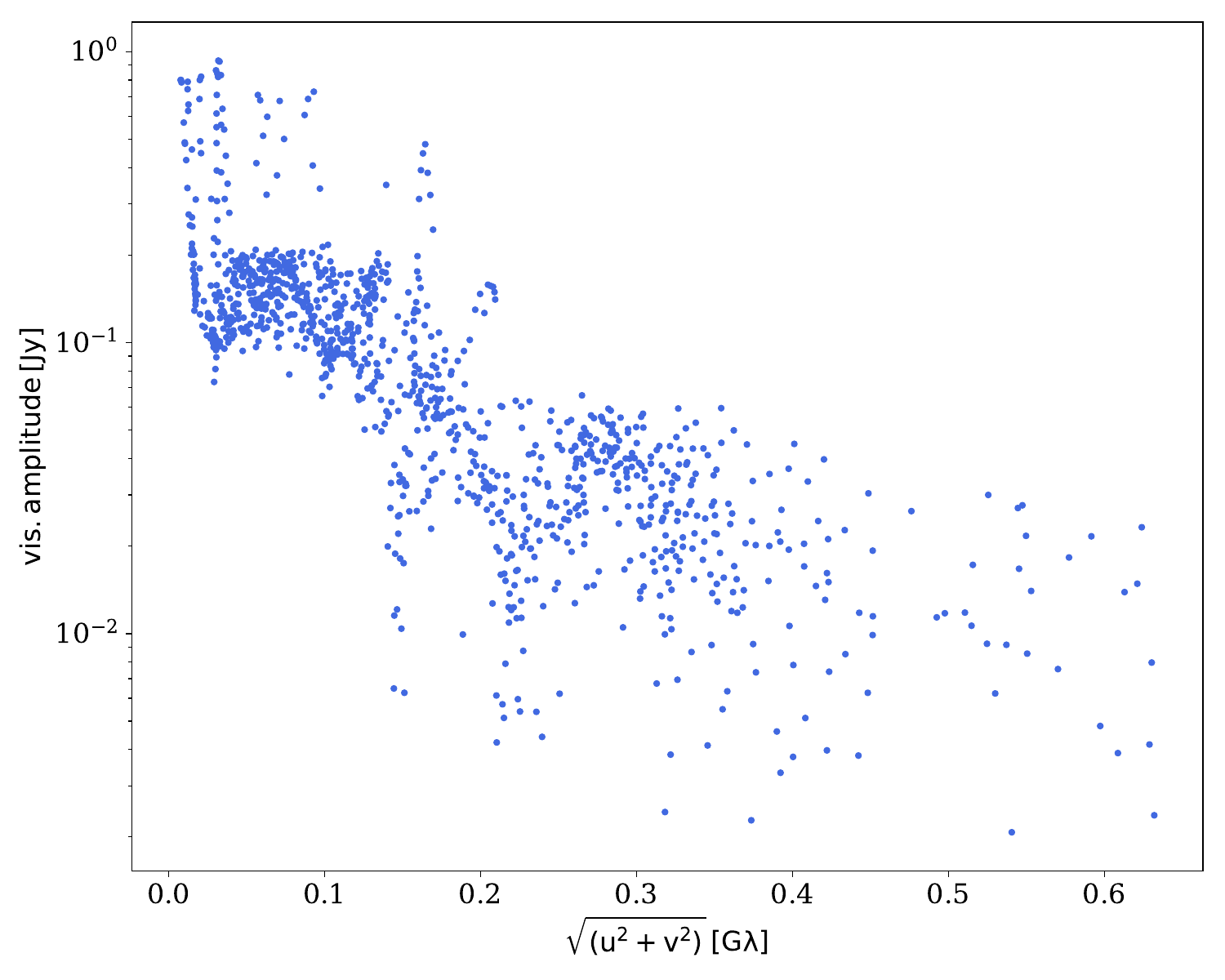}
\caption{Total visibility amplitude plotted against $u-v$ distance for 22~GHz. For clarity, every 2nd data point is plotted with no error bars.}
\label{fig:amplitude}
\end{figure}

During the calculation of $V_{ij}$ we added thermal noise related to the telescope sensitivities, as well as a time dependent gain calibration error extracted from a Gaussian distribution with a standard deviation of 0.05 and a mean of 1 \citep[for details, see][]{2018ApJ...857...23C}. Fig~\ref{fig:amplitude} shows the amplitudes of the final visibilities corresponding to the time step of Fig.~\ref{fig:uv_coverage}, at the central frequency of 22~GHz.

\subsection{Image reconstruction} \label{recons}

The last step to obtain our synthetic emission data was the reconstruction of the synthetic visibilities \citep[for details on image reconstruction in radio astronomy, see][]{2017isra.book.....T}, which we performed with the \texttt{eht-im} code adapted to our data set. In particular, we incorporated the multifrequency functionality described in \citet{2023ApJ...945...40C}. This method simultaneously reconstructs images at different frequencies by parametrizing the intensity as a function of the frequency with a Taylor expansion,

\begin{equation}\label{eq:multifreq}
\log \textrm{I}_i = \log \left[ \textrm{I}_0 \right] + \boldsymbol\alpha \log \left( \frac{\nu_i}{\nu_0} \right) + \boldsymbol\beta \log ^2 \left( \frac{\nu_i}{\nu_0} \right) + \ldots ,
\end{equation}
where $\boldsymbol\alpha$ is the spectral index and $\boldsymbol\beta$ the spectral curvature. As these two parameters are a function of the coordinates, the reconstruction process returns a map of $\boldsymbol\alpha$ and a map of $\boldsymbol\beta$. This multifrequency approach to regularized maximum likelihood (RML) imaging allows the sharing of information across frequencies, acting as a regularizer by improving resolution at lower frequencies and aiding in the recovery of low-brightness structures at higher frequencies. As we can see in Fig.~\ref{fig:spectrum}, the turnover frequency of our spectra is within our range of 8~GHz to 86~GHz. As such, the spectral curvature cannot be neglected, and we include the second-order term on Eq.~\ref{eq:multifreq}.

Before the reconstruction, we added station-dependent systematic noise to improve the convergence of the RML method. We used a Gaussian as a prior image, tilted by 15° as the object \citep[see][]{2016ApJ...829...11C}. At each corresponding time step, this Gaussian had the same flux as the total intensity of the image at the central frequency 22~GHz.

\begin{table}[]
\centering
\caption{Time-averaged values of the major axis, minor axis, and angle of the beam computed from our synthetic observations.}
\label{table:beamparams}
\begin{tabular}{c c c c}
\hline\hline
Frequency  & $b_{maj}$   & $b_{min}$  & $\theta$                                                   \\ \hline
[GHz] & [mas] & [mas] &[$^\circ$] \\ \hline
8   & 2.51 & 1.11 & 178.25 \\ 
15 & 1.34 & 0.59 & 178.25 \\ 
22 & 0.91 & 0.40 & 178.25 \\ 
43 & 0.47 & 0.21 & 178.25 \\ 
86 & 0.34 & 0.07 & 171.90 \\ \hline
\end{tabular}
\end{table}

The reconstruction process returned, for each time step, the maps of $\boldsymbol\alpha$ and $\boldsymbol\beta$ used as parameters in Eq.~\ref{eq:multifreq}. It also returned the output of the RML reconstruction for each of the five frequencies, and that same output convolved with its corresponding synthetic beam. The parameters of said beam associated with each frequency, averaged over every time step, are shown in Table~\ref{table:beamparams}. 

\begin{figure}[h!]
\centering
\includegraphics[width=9cm]{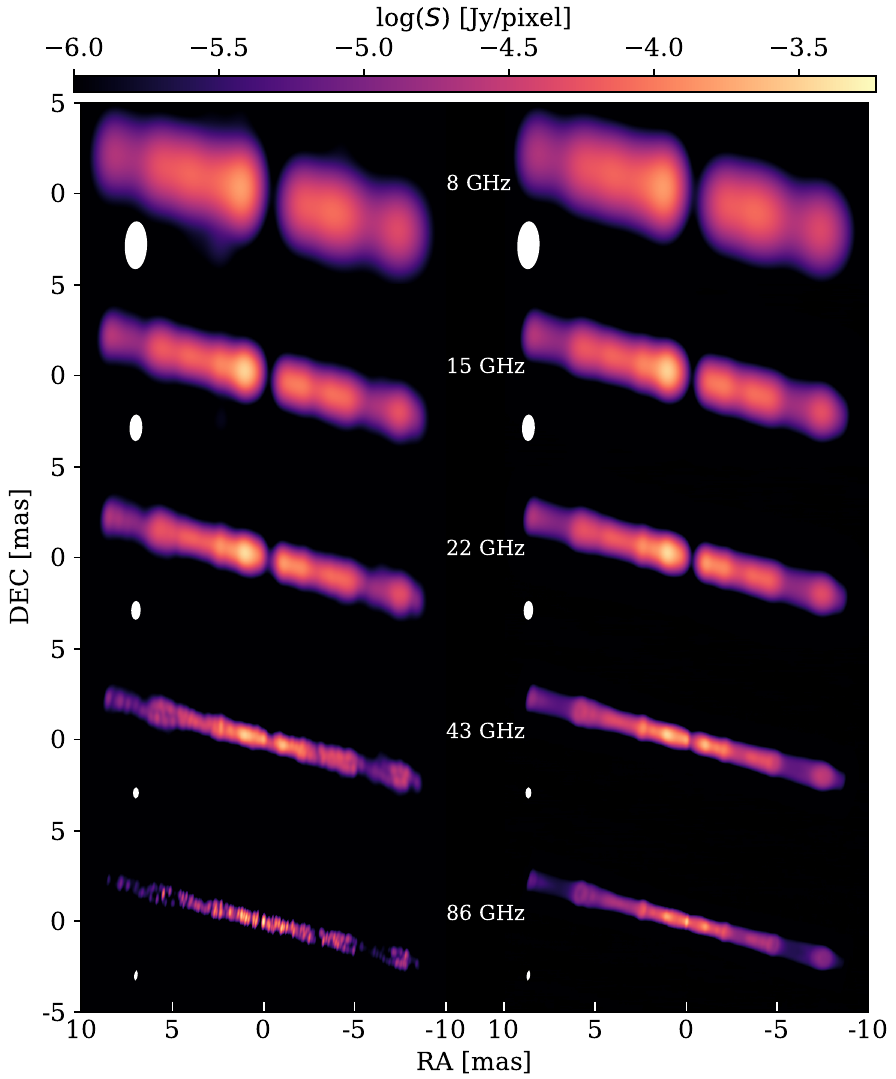}
\caption{Reconstruction of a single time step at five frequencies (left column) contrasted with the direct emission maps (right column). Both have been blurred with the beam of their corresponding frequency, plotted in white.}
\label{fig:theory_vs_mf}
\end{figure}

Figure~\ref{fig:theory_vs_mf} shows the final result of the imaging process for all five frequencies on the initial time step of 2003 October 9. The blurred reconstructions (left) are contrasted with the direct output images blurred with the same beam (right). The frequency dependent emission gap between jets is clearly seen in both the reconstructed images and the blurred direct output. Notice the asymmetry between the left, eastern jet and the right, western counterjet, as well as the locally enhanced emission due to the travelling shocks. The effects of the sparse $u-v$ sampling as well as the limited dynamical range of the observing arrays are best seen at 43\,GHz and 86\,GHz.

\section{Analysis}\label{analysis}

\subsection{Jet kinematics}\label{kinematics}

Identifying bright features, hereafter referred to as components, in jets and tracking their evolution over time allows us to study the jet dynamics including the propagation of shocks and their interactions. Furthermore, by comparing the identified and tracked components with the numerical simulations we can evaluate the impact of the image reconstruction on inferred observed jet dynamics.

As we generated data for 190 time steps ($\sim$ 1.5\,yr) to study over 5 frequencies, we set up an algorithm to find bright features over time as well as to link them together into trajectories. For this, we first used a ``Laplacian of Gaussian''-based method as implemented in scikit-image \citep{scikit-image} \citep[for details on feature finding methods, see][]{1994JApSt..21..225L} on the reconstructed images. These images were previously convolved with a circular beam equal in area to the corresponding elliptical beam for its time step and frequency. On each separate frequency, we then made use of the Trackpy Python package \citep{allan_2024_10696534} over every time step to identify and connect the center of the identified components into trajectories. We removed spurious trajectories from our data set, identified as appearing for only a few frames. These spurious trajectories were mostly present on the 43~GHz and 86~GHz maps due to their noisier reconstructions and smaller convolving beams. We also ignored the components found in the last $\approx$1.7~mas on either side, corresponding to the beam size at 8~GHz. 

\begin{figure}[h!]
\centering
\includegraphics[width=9cm]{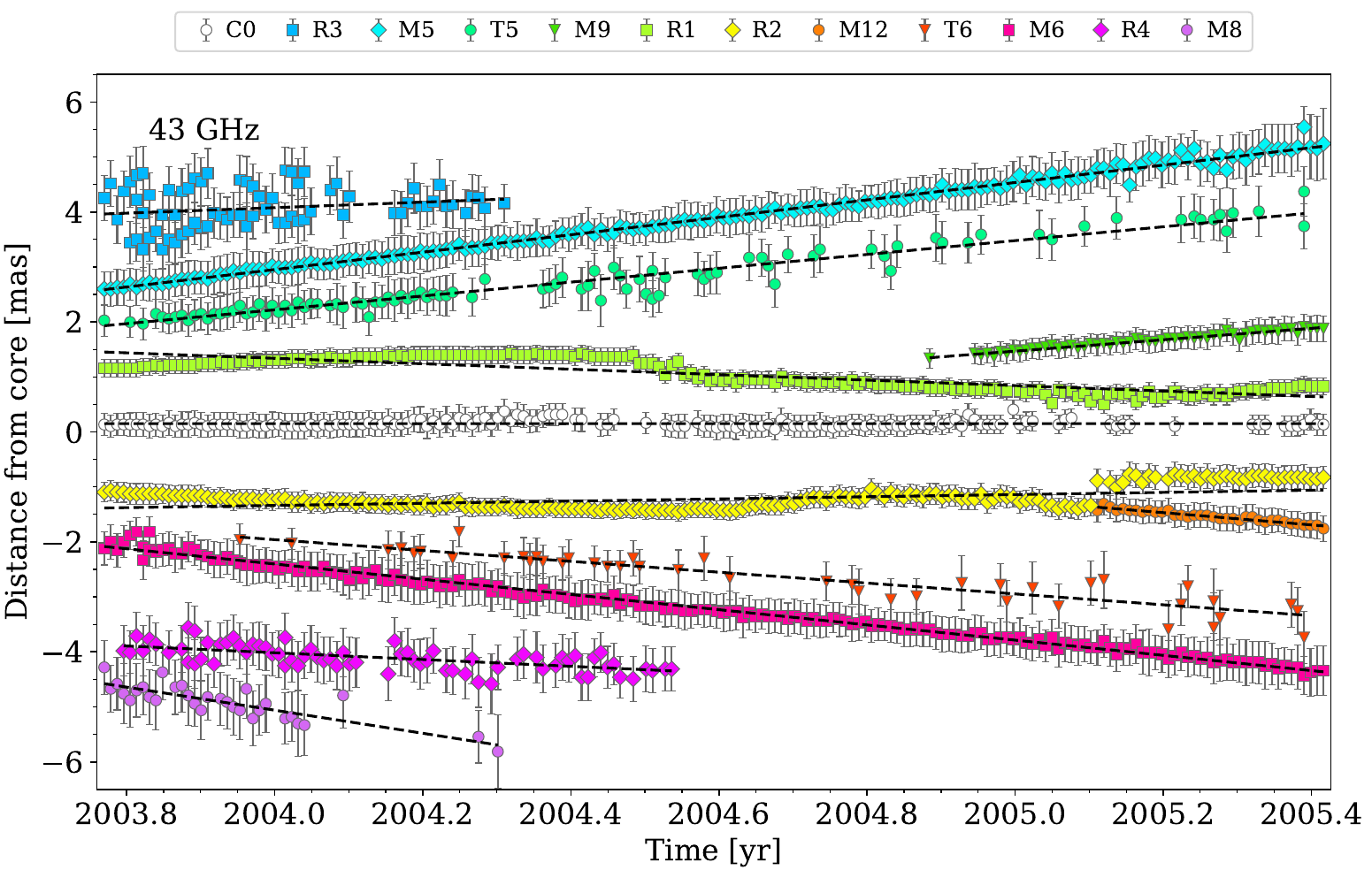}
\caption{Distance of bright components from the central core, identified here as C0, over time for the 43~GHz maps. The components located in the left/right jet are marked with and odd/even number, respectively. Recollimation shocks are indicated by the letter R, moving components by the letter M, their trailing shocks by the letter T followed by the same number of the component which they trail. The dashed lines on each component correspond to a linear fit.}
\label{fig:component43}
\end{figure}

The results for 43~GHz are plotted in Fig.~\ref{fig:component43}, being the frequency that presented the largest amount of detected trajectories, with dashed lines representing linear fits to them. The plots corresponding to the lower frequencies can be found in the Figs.~\ref{fig:component8}, \ref{fig:component15} and \ref{fig:component22}. We contrasted the components across all frequencies and identified them according to their kinematics. The jet core, named C0, is visible in most time steps at this frequency but is obscured by the torus at the lower, optically thick frequencies. Moving components that are travelling outward are identified with the letter M, followed by an odd number for the eastern/left jet (top half of the figures) and an even number for the western/right jet (bottom half of the figures). We furthermore find at this frequency two fainter, harder to detect components moving closely behind M5 and M6. We identify these trajectories as trailing components, named T followed by the number of the moving feature they are trailing with a slower speed. 

The two main recollimation shocks, identified as quasi-stationary components, R1 and R2 are located at roughly 1~mas on either side. While initially staying stationary as expected, they are notably perturbed halfway through our time range. Starting on 2005 in our simulations, two new moving components, M9 and M12, emerge from either recollimation shock. This disturbance is almost entirely gone in the 15~GHz and 22~GHz maps due to the larger beam size blurring adjacent features into each other. In the 8~GHz maps, these travelling perturbations are not detected until they are at least 5~mas away from the core and the recollimation shocks have instead merged with earlier travelling components M5 and M6, shifting both bright features farther out. The secondary recollimation shocks R3 and R4, being fainter than the main R1 and R2, were more difficult to find. However, they were located at roughly 4~mas on either side across most frequencies, particularly visible at 15~GHz and 22~GHz.

\begin{figure}[h!]
\centering
\includegraphics[width=6cm]{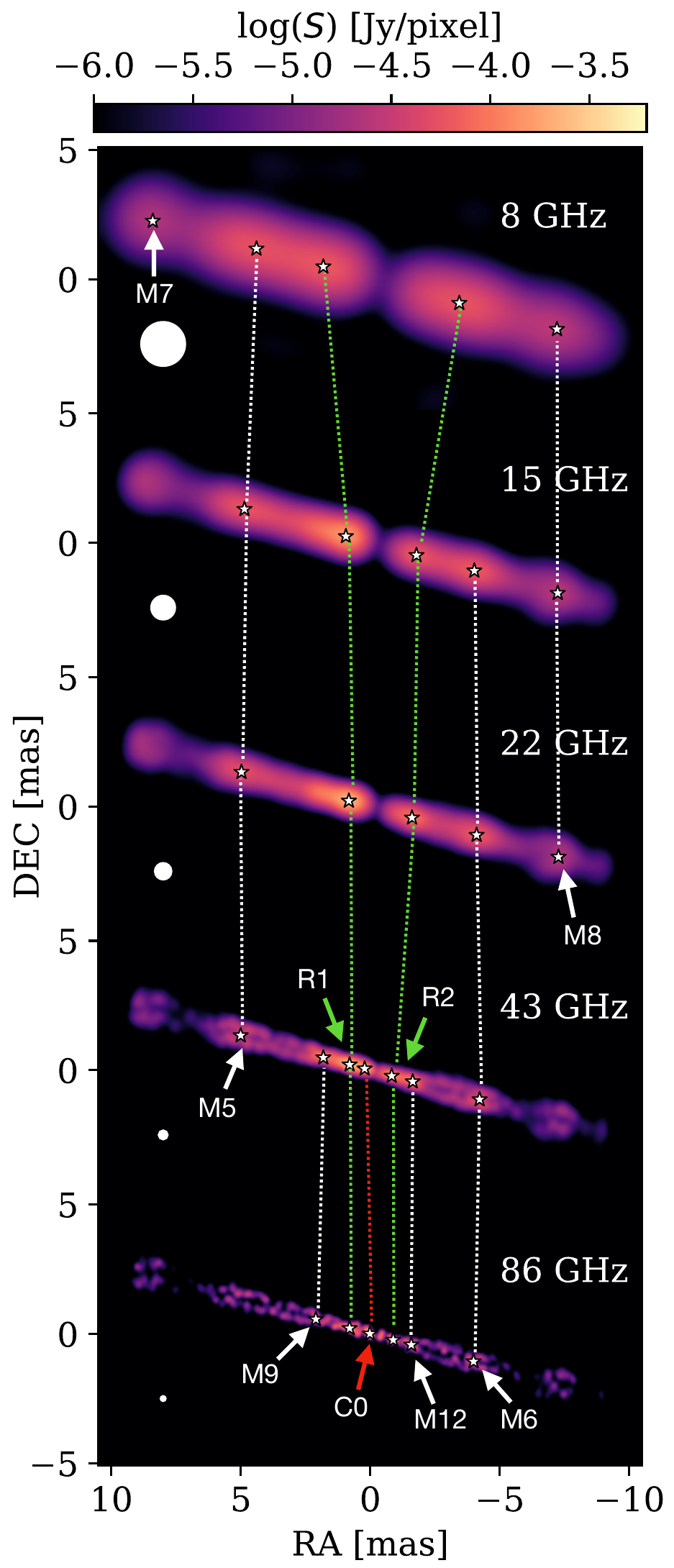}
\caption{Reconstructed and circularly blurred maps at all five frequencies on the 2005 May 5 snapshot. The components detected by our algorithm have been marked. The identified trajectories are connected through the dotted lines, white for the moving shocks, green for the recollimation shocks, and red for the jet core.}
\label{fig:mf_components}
\end{figure}

In Fig.~\ref{fig:mf_components} we show the reconstructions of the 2005 May 5 maps at the five different frequencies, with the central location and cross-identification of the components plotted on top. The primary recollimation shocks R1 and R2, marked with two green lines, show what could be interpreted as an inward displacement between 8~GHz and 15~GHz. However, contrasting the position of the components on Figs.~\ref{fig:component8}, \ref{fig:component15} and \ref{fig:component22} shows that the combined effect of the obscuring torus and the larger resolving beams is the likely cause for this, obscuring inner components and blurring recollimation shocks and nearby travelling components into each other. Fig.~\ref{fig:mf_components} also shows the newly emerging M9 and M12 components, as well as the more distant travelling shocks M5 and M8 becoming too unresolved to be picked up by our algorithm as the frequency increases. As the figure displays one of the final time steps, the fainter secondary recollimation shocks R3 and R4 are not visible, having been respectively obscured by the passing shocks M5 and M6. While Fig.~\ref{fig:initemission} shows three maxima in flux on either jet, the missing ones at roughly 2.5~mas may be obscured by the brightness of the moving components and of the nearby brighter shocks at 1~mas.

\begin{table*}[]
\centering
\caption{Fitted velocities to the moving components.}
\label{table:velocities}
\begin{tabular}{c|ccccc}
\multicolumn{1}{l|}{} & \multicolumn{5}{c}{$\textrm{v}_{\textrm{app}}$ {[}c{]}}         \\ \hline
ID                    & 8 GHz & 15 GHz & 22 GHz & 43 GHz & 86 GHz \\ \hline
C0                    & -     & -      & -      & $0.0001\pm0.0116$ & $0.013\pm0.005$  \\
R1                    & $0.29\pm0.04$  & $-0.05\pm0.02$ & $-0.11\pm0.02$  & $-0.164\pm0.008$  & $-0.132\pm0.009$  \\
R3                   & $0.21\pm1.46$     & $0.13\pm0.65$      & $0.04\pm0.41$      & $0.17\pm0.11$  & - \\
M5                    & $0.73\pm1.48$     & $0.49\pm0.04$   & $0.52\pm0.03$   & $0.52\pm0.02$   & $0.49\pm0.03$      \\
T5                    & -     & -      & -      & $0.42\pm0.03$   & -      \\
M7*                  & $1.26\pm0.30$  & $0.76\pm0.10$   & $0.72\pm0.08$   & -      & -      \\
M9                   & -     & -      & -      & $0.34\pm0.06$   & $0.47\pm0.08$   \\
R2                    & $0.06\pm0.05$ & $-0.01\pm0.03$ & $-0.05\pm0.02$  & $0.066\pm0.010$   & $0.060\pm0.008$   \\
R4                   & -     & $-0.12\pm0.18$      &  $-0.099\pm0.103$      & $-0.20\pm0.07$  & - \\
M6                    & - & $-0.47\pm0.11$  & $-0.43\pm0.04$  & $-0.46\pm0.02$  & $-0.47\pm0.02$      \\
T6                    & -     & -      & -  & $-0.33\pm0.05$  & -      \\
M8                  & $-0.91\pm0.55$ & $-0.69\pm0.12$  & $-0.67\pm0.08$  & $-0.70\pm0.25$      & -      \\
M10*                  & $-0.44\pm0.64$ & $-0.47\pm0.43$  & $-0.48\pm0.33$  & -      & -      \\
M12                   & -     & -      & -      & $-0.37\pm0.13$  & $-0.37\pm0.08$
\end{tabular}
\tablefoot{
	The components marked with an asterisk indicate that the velocity was computed in the limited time range where the component was determined to be moving with constant velocity.}
\end{table*}

Table~\ref{table:velocities} shows the slope of the trajectories plotted in Fig.~\ref{fig:component43} with a dashed black line, as well as the trajectories for the other frequencies. As a component reaches the end of the simulated box, it cannot be tracked any further. This gives the impression that the components slow down as they reach the end of the simulated region. This is a limitation of the hydrodynamic simulation, and not a physical phenomenon. We marked with an asterisk the velocities that were computed up to an intermediate time step, determined from the residuals of a linear fitting.

As expected, the jet core C0 only appears at the optically thin frequencies with a velocity compatible with zero. The recollimation shocks generally have values oscillating around zero, although perturbed by moving shocks and the limited resolution. For example, the faster velocity of R1 in 8~GHz is accompanied by an anomalously high velocity for M5 when compared to all four of the other frequencies. This disagreement is present in most trajectories visible at 8~GHz, most notably M7 and M8. M7, which shows the biggest disagreement, is difficult to pick up at this frequency until it travels enough to be far away from the faint recollimation shock R3. This is due to the combined effects of the large blurring beam and the obscuring torus, and it makes 8~GHz a nonoptimal frequency to study jet dynamics at this scale.

The travelling component M6 shows a largely consistent behavior across frequencies, as well as the late-appearing M12. M9 shows a larger disagreement between 43~GHz and 86~GHz data. The velocities computed for M5, M7, and M8 are also in agreement when 8~GHz data is excluded. We thus match them to the perturbations injected at the nozzle in the hydrodynamic simulation with a velocity of $0.5\,c$. 

\subsubsection{Positional errors}

As previously mentioned, we used an automated feature-finding algorithm to locate the components in our map. This was done due to the large number of images, 190 time steps over 5 frequencies. In observational studies, the component modeling and cross-correlation are typically done on the interferometric $u-v$ plane \citep[see][]{2009AJ....137.3718L}. As we performed our analysis on the image plane, we could not calculate positional errors based on the component fitting results as described in \citet{2009AJ....137.3718L}. 

We assigned the location of each component an error, with the criteria being that it should depend on both the observing frequency and the flux density of the component. For the former, we related the error to half of the full width at half maximum (FWHM) of the circular blurring beam at each frequency and time step, $b/2$. For the latter, we assigned a weight to each component, $w_i=S_i/S_{\textrm{max}}$. $S_i$ is the flux density at the center of a given component, while $S_{\textrm{max}}$ is the maximum flux density found in the map at that frequency and time. As can be seen in Fig.~\ref{fig:initemission}, this tracks both the presence of shocks and the distance to the core, as the latter relates to the underlying flux profile. 

By combining these two criteria, and following the relation between weights and variance $w_i=1/\sigma_i^2$, we assign to each of our components the positional error

\begin{equation}\label{eq:error}
\delta x_{i} = \sqrt{\frac{S_{\textrm{max}}}{S_i}}\frac{b}{2}.
\end{equation}

This equation produces a minimum error of $b/2$, larger than the minimum error of $b/10$ adopted by \citet{2019A&A...623A..27B}. This more conservative approach is warranted to ensure that adjacent components identified by the automated feature-finding algorithm in the image plane are truly distinct. It can be seen in Fig.~\ref{fig:component43}, and especially in Fig.~\ref{fig:component86}, that for the higher frequencies this error displays more time variability. This is caused by the low flux density at these frequencies affecting the quality of the reconstructions.

\subsubsection{Shock-shock interaction}

\begin{figure}[h]
\centering
\includegraphics[width=9cm]{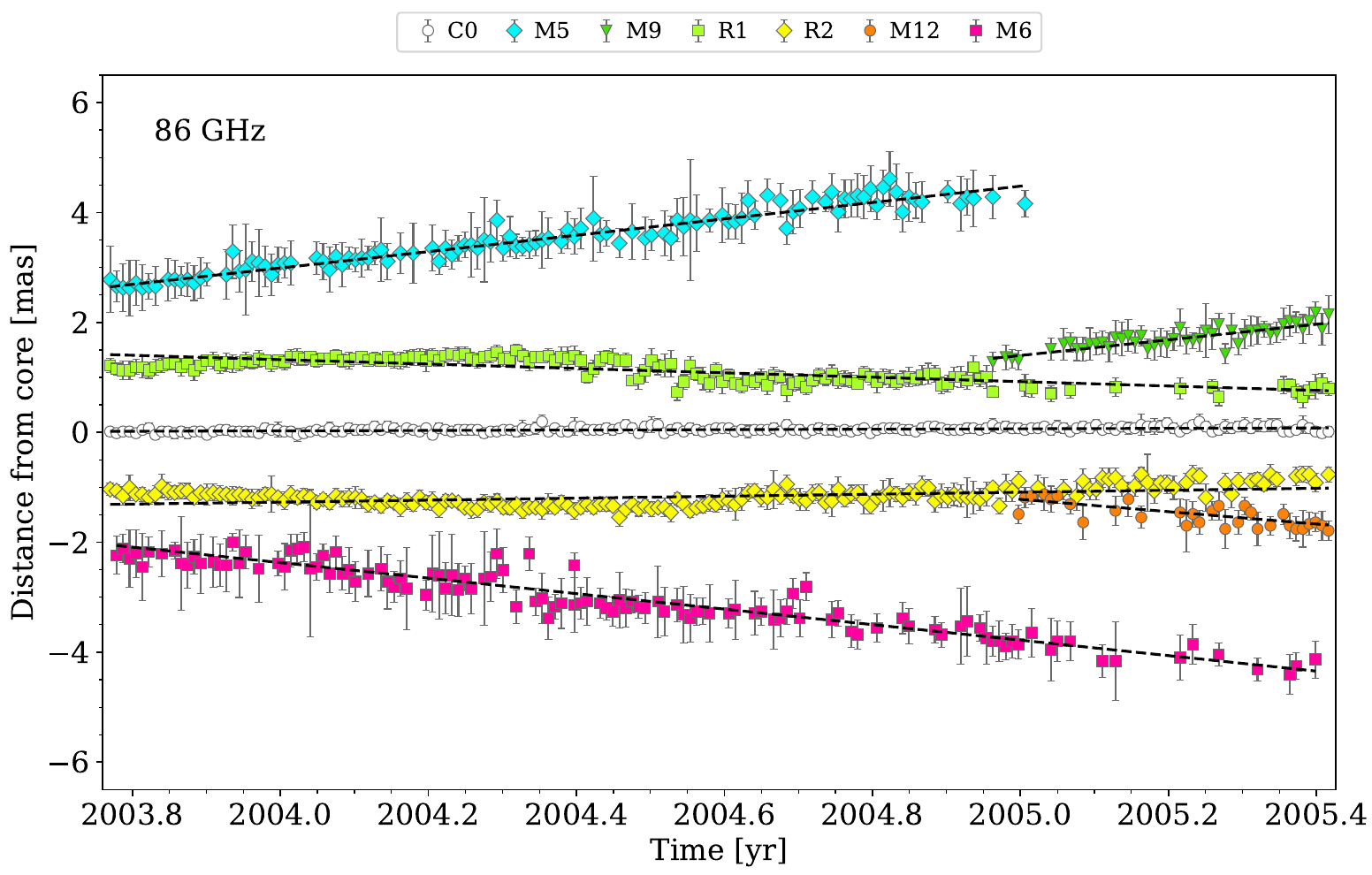}
\caption{Distance of bright components from the central core over time for the 86~GHz maps. The components are marked by a letter and a number as explained in Fig.~\ref{fig:component43}. The dashed lines on each component correspond to a linear fit.}
\label{fig:component86}
\end{figure}

The maps at 86~GHz contain the lowest flux, which hinders the reconstruction process. This strong impact of the reconstruction process on the data is seen in Fig.~\ref{fig:timespaceplot}. While this means that outer components cannot be properly detected, the low beam size at this frequency does allow us a closer look into the innermost areas of the jet. In Fig.~\ref{fig:component86} we have plotted the distance of the components from the core for 86~GHz, where it can be observed that features become too unresolved for detection at roughly 4~mas. The trajectories that can be found at this frequency are consistent with those seen in Fig.~\ref{fig:component43} at 43~GHz. 

\begin{figure}[h!]
\centering
\includegraphics[width=8cm]{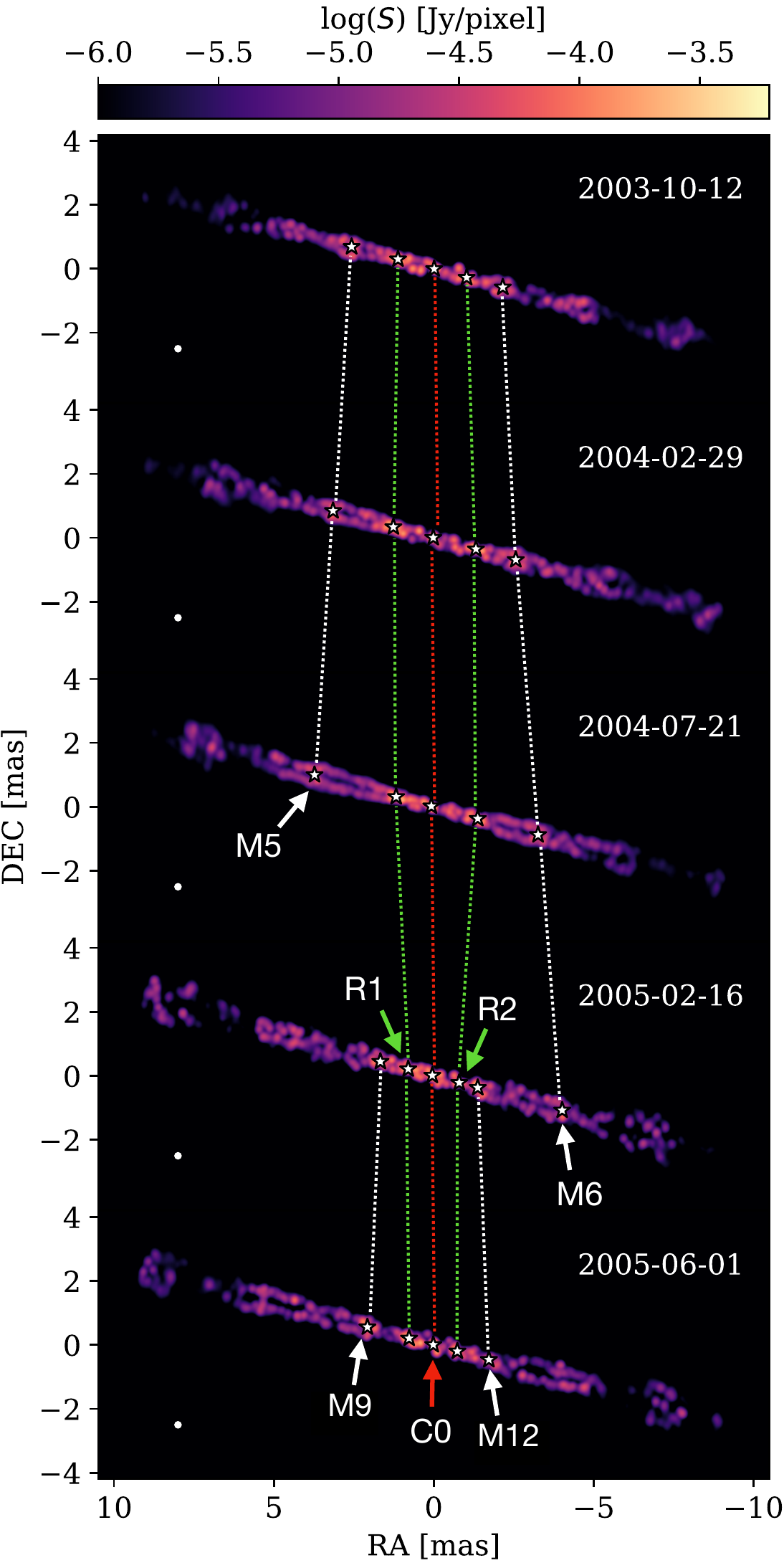}
\caption{Five different snapshots of reconstructed and circularly blurred 86~GHz maps. The components detected by our algorithm have been marked. The identified trajectories are connected through the dotted lines, white for the moving shocks, green for the recollimation shocks, red for the jet core.}
\label{fig:componentmultistep}
\end{figure}

The resolution at these two frequencies allows us to see the two newly emerging components, M9 and M12, as well as an oscillatory movement of the recollimation shocks. Using the axial emission of the initial steady-state model at 43~GHz (see Fig.~\ref{fig:initemission}) as a guide, we would expect three recollimation shocks on either jet. R3 and R4 can be identified with the peaks between 4~mas and 5~mas, but they disappear early on in our simulation due to the passing of travelling shocks. R1 and R2 appear at all frequencies and most time steps, but seem to travel away from and back to their initial position when crossed by new components. Fig.~\ref{fig:componentmultistep} shows five different time steps of the reconstructed 86~GHz maps with highlighted components. The last two time steps show M9 and M12 emerging as well as, as mentioned, R1 and R2 moving inwards. We will later discuss the possible reasons for the behavior of the standing shocks, and the impact of moving shocks on jet structure.

\subsubsection{Viewing angle and intrinsic speed estimation}

The apparent speeds of travelling components allow us to probe the symmetry 
in an independent way. Assuming that the jets are intrinsically symmetric, the ratio $R$ of total flux density between the approaching (left) and the receding (right) jets is described by the equation:

\begin{equation}
	 R = \frac{S_{\textrm{a}}}{S_{\textrm{r}}} = \left( \frac{1+\beta\cos\theta}{1-\beta\cos\theta} \right)^{2-\alpha} 
	\label{eq:flux_ratio} .
\end{equation}

$\beta$ is the intrinsic jet speed in units of $c$, $\theta$ is the viewing angle, and $\alpha$ is set to $-1$ following \citet{2019A&A...623A..27B}. Furthermore, the apparent speeds of Table~\ref{table:velocities} relate to the intrinsic jet speed and viewing angle as:

\begin{equation}
	 \beta_{\textrm{app,a/r}} = \frac{\beta\sin\theta}{1\mp\beta\cos\theta}
	\label{eq:app_speed} .
\end{equation}

\begin{figure}[h!]
	\centering
	\includegraphics[width=9cm]{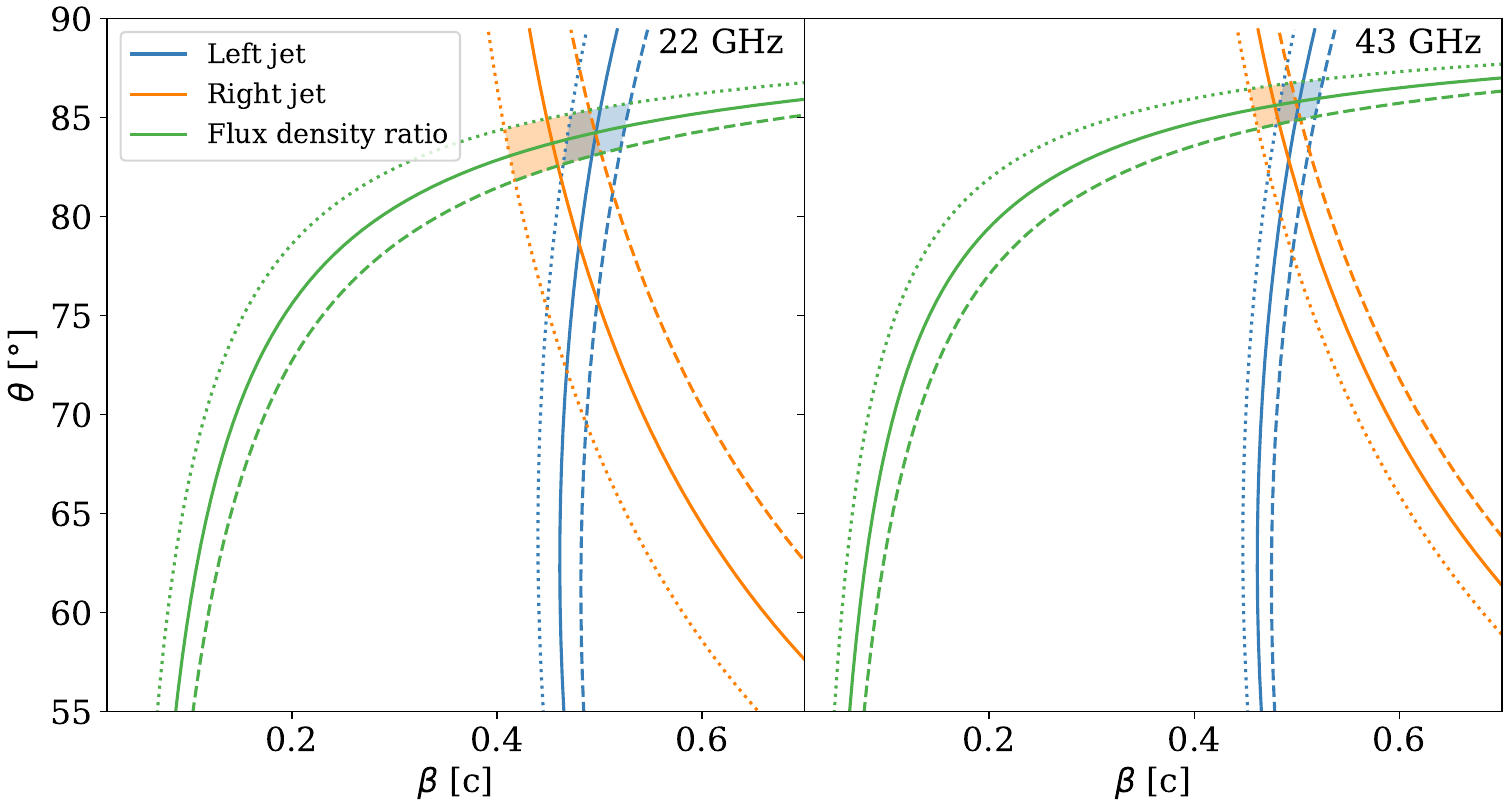}
	\caption{Allowed parameter space for the intrinsic speed $\beta$ and viewing angle $\theta$ for both jets, at 22~GHz (left) and 43~GHz (right). The orange and blue lines represent Eq.~\ref{eq:app_speed}, the green lines Eq.~\ref{eq:flux_ratio}. The dotted and dashed lines represent the lower and upper error range. Both cases allow intrinsic symmetry, as shown by the overlapping shaded areas.}
	\label{fig:parameter_space}
\end{figure}

The overlap of Eq.~\ref{eq:app_speed} with Eq.~\ref{eq:flux_ratio} defines the allowed parameter space for $\beta$ and $\theta$. If symmetry holds, the two parameter spaces created by both signs of Eq.~\ref{eq:app_speed} should furthermore overlap with each other, within the derived uncertainties. Fig.~\ref{fig:parameter_space} shows the result of the analysis for our reconstructed data at the central frequencies of 22~GHz and 43~GHz. We used the time-averaged mean of $R$ and its standard deviation, while the values for $\beta_{\textrm{app}}$ were taken from Table~\ref{table:velocities}. We chose the components M5 and M6, as they were the main travelling shocks during our time range and were picked up with minimal scatter for most frequencies. These properties indicate they are likely travelling shocks and thus faster than the underlying flow, and so the following results should be taken as an approximate limit.
	
It can be seen that both cases allow for intrinsic jet symmetry, even with the more restrictive errors of 43~GHz. Furthermore, both parameter spaces contain the ground truth of $\beta = 0.5$ that was set into our simulations. The viewing angle of 80$^\circ$ is overestimated by roughly 5$^\circ$. As the allowed viewing angles are smaller for 22~GHz, an optically thicker frequency, it's possible that our flux ratio $R$ was overestimated due to the core emission. Still, this result shows that an intrinsically symmetric jet could be retrieved from the data even after undergoing a reconstruction process.

\subsection{Jet collimation profiles}\label{collimation}

\begin{figure*}[h!]
\centering
\includegraphics[width=17cm]{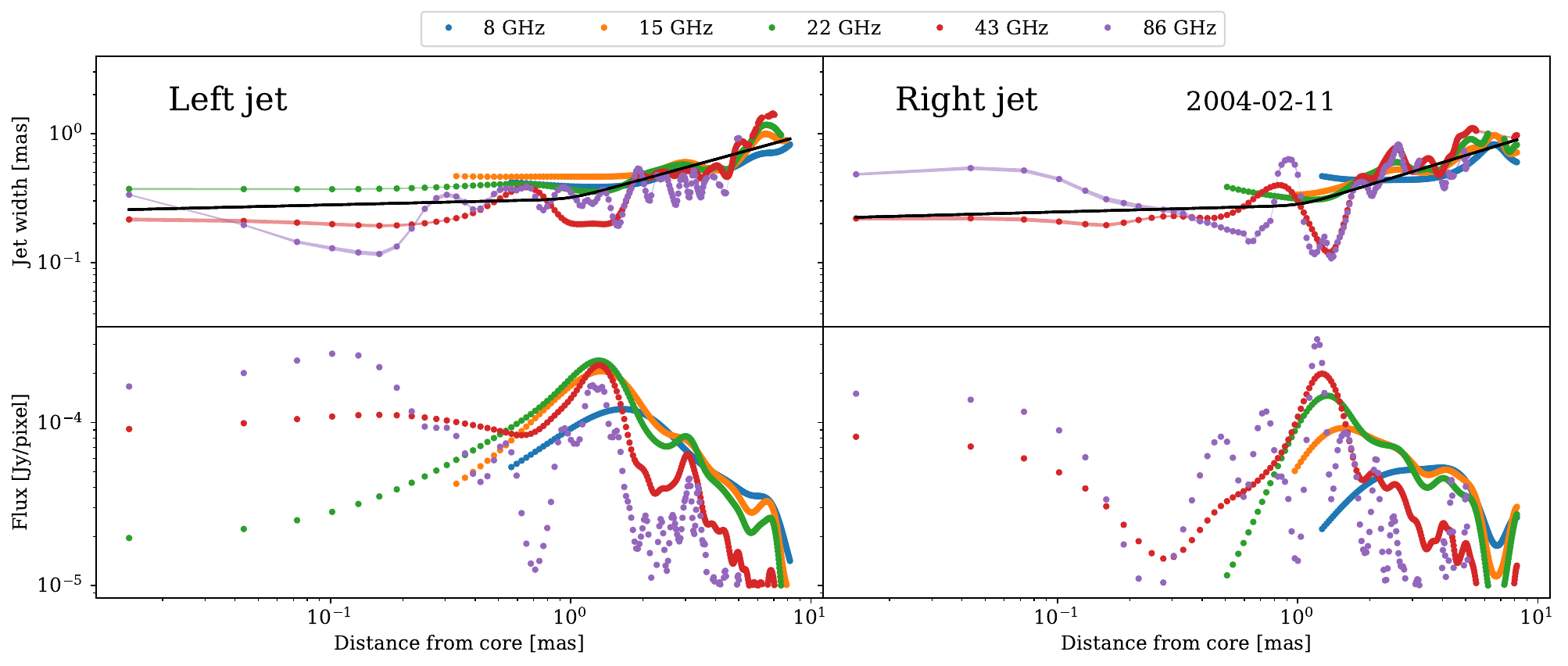}
\caption{Width and peak flux profiles for the left (eastern) and right (western) jet at a single time step. The black solid line represents the broken power-law resulting from fitting all frequencies together.}
\label{fig:collimation_profile}
\end{figure*}

In this section we analyze the profiles of the width, i.e., jet collimation profiles, of both jet and counterjet, using the following method. 

As in the previous section, the flux maps resulting from the imaging process were blurred with a circular beam equal in area to the corresponding elliptical beam for its time step and frequency. The image was then rotated back by 15$^{\circ}$ so that the jet spine was aligned with the x-axis, for ease of computing. From there, the peak flux and jet width were evaluated at every slice perpendicular to the jet spine. To compute the jet width, a Gaussian fit was performed on each slice, from which we obtained a FWHM equivalent to the width of the blurred jet. A flux-dependent error was incorporated into the Gaussian fit. This error was computed from taking the mean value of an empty region of the flux map to obtain an average value of the absolute error, and dividing it by the map to obtain a map of relative errors.

The final jet width was calculated by deconvolving the circular blurring beam and the width of the Gaussian fit, obtaining a jet width of $\sqrt{\text{FWHM}_{\text{fit}}^2-\text{FWHM}_{\text{beam}}^2}$. This was done over each time step of every frequency, and these deconvolved values were the ones we used for the purpose of studying jet collimation. 

Following the method used for the study of the collimation of the NGC~1052 jets in \citet{2022A&A...658A.119B}, a broken power-law model was fit to the data at all frequencies simultaneously,

\begin{equation}
w(d) = W_0\, 2^{\left(k_\textrm{u}-k_\textrm{d}\right)/ s} \left(\frac{d}{d_\textrm{b}}\right)^{k_\textrm{u}}\left[1+\left(\frac{d}{d_\textrm{b}}\right)^s\right]^{\left(k_\textrm{d}-k_\textrm{u}\right)/s},
\label{eq:broken_power_law}
\end{equation}
where w is the jet width as a function of the distance to the core d and $d_\textrm{b}$ is the breakpoint between the upstream and downstream sections of the jet, defined respectively as the region between the core and the breakpoint, and the region beyond the breakpoint. $\textrm{W}_0$ is the width at the breakpoint, and $\textrm{k}_{\textrm{u}/\textrm{d}}$ are the power-law slopes of the upstream and downstream sections respectively. s is a free parameter which controls the smoothness of the transition between power laws, set to 10 following \citet{2020AJ....159...14N} and \citet{2022A&A...658A.119B}. The position of the breakpoint is expected to correspond on each side to the primary recollimation shock, an area characterized by its narrow width and high flux. A recollimation shock is present at approximately 1~mas from the core at all time steps, pointed out by green lines in Fig.~\ref{fig:timeevo}. These can be seen in the reconstructed images in Fig.~\ref{fig:theory_vs_mf}, particularly at 43~GHz where there is no significant obscuring from the torus. It would be expected for $d_\textrm{b}$ in Eq.~\ref{eq:broken_power_law} to match these main recollimation shocks, which mark the transition from upstream to downstream jet. While our algorithm would have an easier time positioning this $d_\textrm{b}$ on our direct emission maps, it is of interest for this paper to examine the ways in which this analysis was affected by the noise and limited resolution of a reconstruction process. Figure~\ref{fig:collimation_profile} shows the result of the broken power-law fitting for one time step. This fitting was done simultaneously for all frequencies, performed separately over the left jet and right counterjet after removing the invalid values of low-flux regions. We focus our analysis on the values of the downstream slope, $k_\textrm{d}$, for which we obtained single epoch relative errors on the order of 0.1\% for most time steps.

\begin{figure}[h]
\centering
\includegraphics[width=9cm]{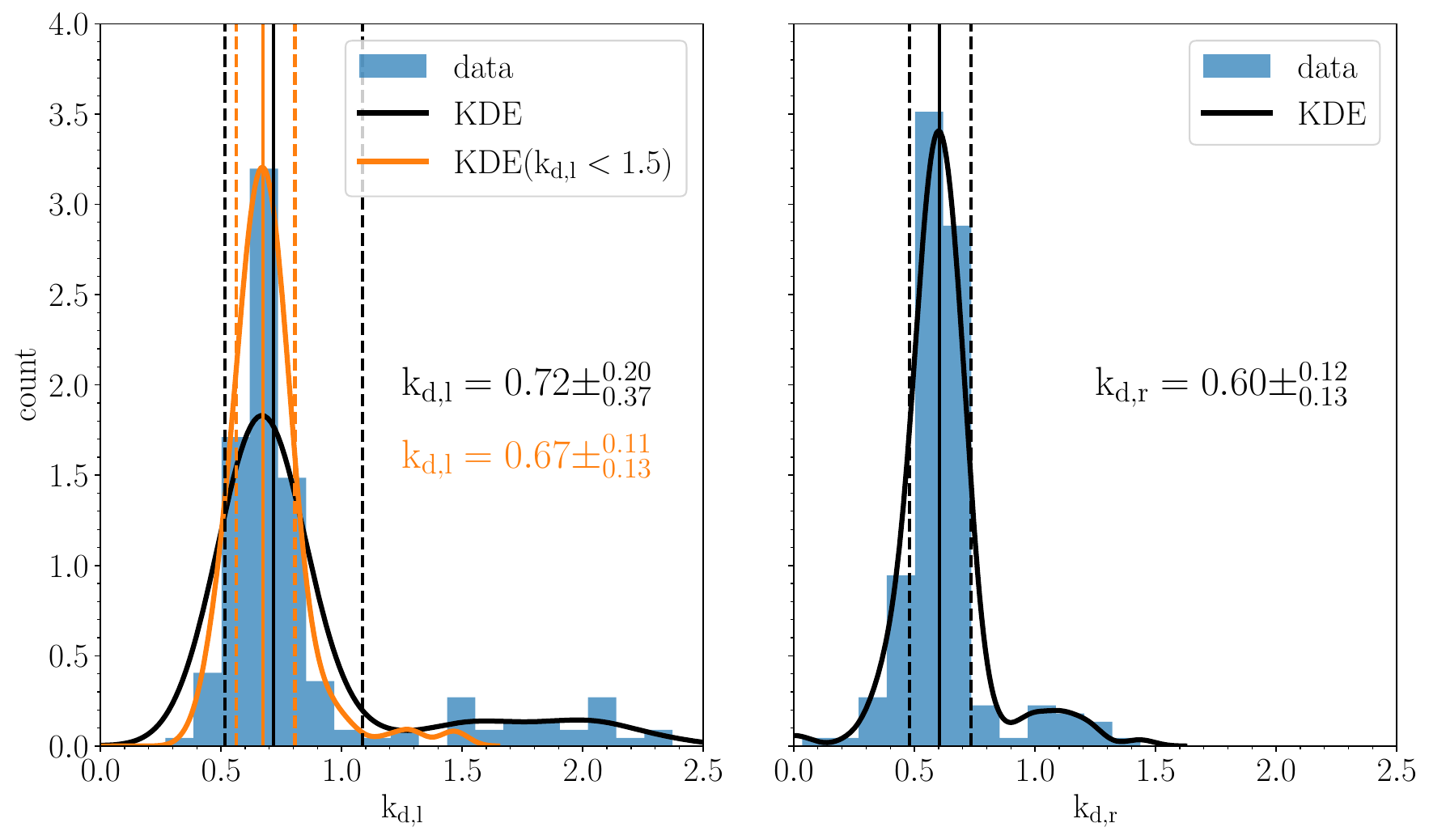}
\caption{Kernel density estimation of the distribution of downstream slopes resulting from the broken power-law fitting.}
\label{fig:kde}
\end{figure}

 We performed a kernel density estimation (KDE) analysis on the $k_\textrm{d}$ distribution of both jets for all time steps. This allowed us to estimate a probability density function and to draw the values of the median as well as the quartiles of the distribution. The results are plotted in Fig.~\ref{fig:kde}, where the solid black line drawn over the histogram is the KDE result. We show the median as well as the first and third quartiles, which we used as a range. The resulting $k_\textrm{d,l}=0.72^{+0.20}_{-0.37}$ for the left jet and $k_\textrm{d,r}=0.60^{+0.12}_{-0.13}$ for the right counterjet point toward the jet and counterjet not being significantly asymmetric on a global scale. As a test, we ran the analysis again, removing $k_{\textrm{d,l}}>1.5$ values from our data set. These abnormally high values of the downstream slope could correspond to a failure of the fitting algorithm to identify the position of the recollimation shock as the breakpoint of the broken power law. The new analysis constrained the median value further, $k_\textrm{d,l}=0.67^{+0.11}_{-0.13}$. This is seen as the orange line on the left panel of Fig.~\ref{fig:kde}.
 
  A more detailed study of the validity of these fittings is found in the Appendix~\ref{appendixfitting}. Carrying out this analysis on the direct emission maps supports our conclusion that the tail present in Fig.~\ref{fig:kde} is an artifact introduced by failed fittings on reconstructed images. Furthermore, the range provided by our KDE calculation narrows by a factor of $\sim$2, while the new values of $k_\textrm{d,l}=0.69^{+0.05}_{-0.11}$ and $k_\textrm{d,r}=0.66^{+0.14}_{-0.07}$ remain statistically equivalent to those of Fig.~\ref{fig:kde}. While it is of interest for this study to focus on the impact of the reconstruction process on the study of jet kinematics, this test is in agreement with the intrinsically symmetric nature of our simulated observation.

\begin{figure}[h]
\centering
\includegraphics[width=8cm]{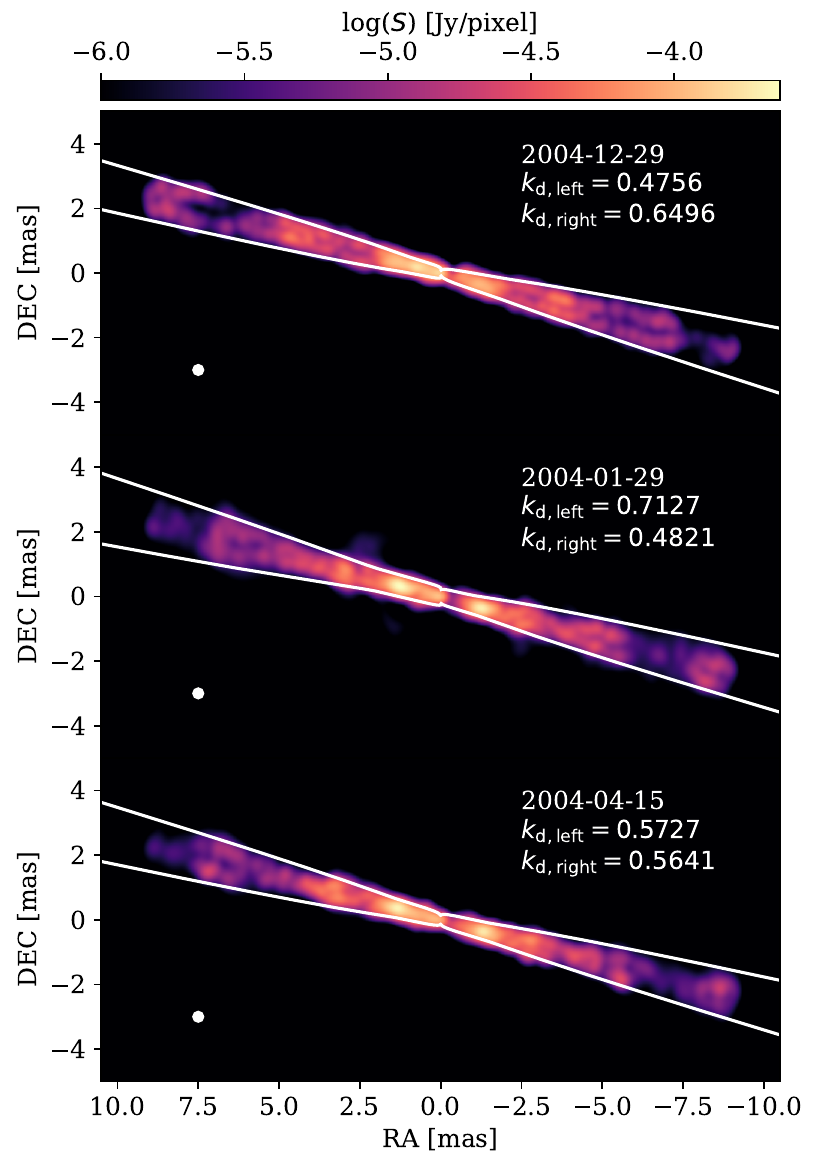}
\caption{Emission maps at 43~GHz with their resulting broken power-law fittings in solid white lines, as well as the values for the downstream slopes for eastern and western jets.}
\label{fig:threeslopes}
\end{figure}

 Still, downstream profiles with a significant asymmetry were found when looking at specific time steps. Figure~\ref{fig:threeslopes} shows a snapshot where the downstream jet has a wider collimation profile than the counterjet and vice versa, as well as a case where they are roughly symmetric. This shows that while an asymmetry may arise at specific time steps due to light aberration and the shock structure of the jet caused by a travelling perturbation, a significant asymmetry cannot be claimed on the global profiles over all epochs.

\section{Discussion}\label{discussion}

In this paper, we present new results from numerical simulations to which we have incorporated the aberration caused by the finite speed of light. Past studies showed that relativistic beaming may not be enough to properly explain the dynamics of apparent superluminal motion in AGN jets, making light aberration absolutely necessary. Not accounting for these slow-light calculations may lead to an underestimation of the flux of the moving components, and it may also affect the overall profile of the jet. Concerning the flux, following the analysis of  \citet{1997MNRAS.288..833K} with our values for the viewing angle and injection velocity, 80$^\circ$ and $0.5\,c$, we calculate that without slow-light we would have underestimated the flux of the moving components by around 10~\%. This is less dramatic than the example given of a typical source with apparent superluminal motion, where the flux could be underestimated by an order of magnitude. This is expected for two-sided sources observed at subluminal speeds such as NGC~1052, whose large viewing angle makes it ideal for observational comparisons. Still, the introduction of light aberrations affects not only the observed brightness of the moving components, but also their size and shape. Figure~\ref{fig:fastslow} shows how accounting for light travel time not only introduces a bipolar asymmetry, but also alters the collimation profile of the jets due to the rotation and broadening of the shocks.

We also introduced a synthetic imaging method where several VLBI frequencies are reconstructed simultaneously, allowing the sharing of information between the different frequencies. \citet{2023ApJ...945...40C} introduce the method and show the results of computing a map of the spectral index during the reconstruction. They also show the necessity of using nonzero values of the spectral curvature in frequency ranges where the curvature is not negligible, as is the case here. As we covered frequencies from 8~GHz to 86~GHz, this significantly improved our reconstruction.

During our component tracking, we found two trailing shocks, T5 and T6. These were only found in the 43~GHz maps, where the noise in the reconstruction is much larger to the point where the detection threshold had to be increased, and thus could be spurious features. To better understand to which extent these features are affected by the reconstruction process, Fig.~\ref{fig:timespaceplot} shows our trajectories on top of a map where the one-dimensional flux of the jet spine is plotted against time. This is done for the reconstructed images as well as for the blurred direct maps. It does appear visually as if trailing shocks could be accompanying the moving components. This is particularly visible for components M7, M5, and M6. However, it appears as though our tracking algorithm in its current settings has difficulties in separating them from the main components. The trailing component T5 could, however, be retrieved near the beginning and the end of the direct maps. 

\citet{Agudo2001} and \citet{2009ApJ...696.1142M} carried out SRHD simulations which showed that perturbations travelling along a jet can lead to the formation of trailing shocks emerging behind a sufficiently strong moving component. They furthermore concluded that these trailing components should move slower than the main feature which they are trailing. In addition, \citet{2022A&A...661A..54F} identify these trailing components with relaxation shocks that arise from the interaction of a moving shock passing through a standing shock. The findings that the trailing shocks follow the main components with a slower velocity are consistent with the velocities we calculated for T5 and T6. These works furthermore study the oscillation of standing components after being perturbed by a passing shock. This effect is visible in our results, particularly for the higher-resolution 43~GHz and 86~GHz maps. Rather than staying at the same distance on either side, the recollimation shocks appear to move inwards halfway through our simulation. For our selected time range, the recollimation shocks are disturbed by the passing of the shocks M5 and M6, before the starting point of the data set shown in this work. Starting roughly halfway, these recollimation shocks slide inwards, and a possible outward motion is starting to appear near the end, caused by the newly emerging shocks M9 and M12. This is displayed in Fig.~\ref{fig:timespaceplot}.

Several recent works study the symmetry of NGC~1052, making it a good source to compare our results to. These studies fit a broken power-law to the width profile of the jet, yet there is no consensus on whether an asymmetry is present. Using 2000-2001 observations, \citet{2020AJ....159...14N} calculate the upstream values of the slope to be consistent with zero for both jet and counterjet, and a downstream slope value of $0.96\pm0.04$ for the eastern jet and $1.02\pm0.03$ for the western jet, reaching the conclusion of symmetry. On the other hand, using 2017 observations, \citet{2022A&A...658A.119B} calculate $1.01\pm0.01$ downstream for the eastern jet and jet $1.22\pm0.02$ for the western jet, with an upstream value of $0.17$ for both. They carry out the same analysis on stacked VLBA observations from 2005 to 2009, calculating downstream values of $0.80\pm0.01$ for the eastern jet and $1.22\pm0.05$ for the western jet, and upstream values of around $0.20$ for both. \citet{2020MNRAS.495.3576K} only calculate the slope values for the eastern jet, obtaining a steeper value of $0.391\pm0.048$ for the upstream jet, and $1.052\pm0.081$ downstream. 

By comparison, carrying out the same broken power-law fitting on our reconstructed images yielded upstream slopes of $k_{\textrm{u}}\leq 0.5$ for the majority of time steps. 93\% and 85\% of the time steps showed $k_{\textrm{u}}\leq 0.25$ for the left and right jet, respectively, while this percentage rose to 100\% when repeating this analysis on the infinite resolution images. This collimation profile between cylindrical and parabolic ($0\leq k \leq 0.5$) is consistent with recent studies. Furthermore, the final analysis of our reconstructed images yielded downstream slopes of $k_\textrm{d,l}=0.67^{+0.11}_{-0.13}$ for the left jet and $k_\textrm{d,r}=0.60^{+0.12}_{-0.13}$ for the right counterjet. Testing our analysis on the direct emission maps narrowed the range further to $k_\textrm{d,l}=0.69^{+0.05}_{-0.11}$ and $k_\textrm{d,r}=0.66^{+0.04}_{-0.07}$. While this is significantly less conical than VLBI observations of NGC~1052, the underlying collimation profile is determined by the chosen pressure gradient of the ambient medium. The aim of this work was to determine whether it was possible to induce asymmetries to the extent found in \citet{2022A&A...658A.119B}, when starting from an intrinsically symmetric double-sided jet. Our results point to the conclusion that perturbing the jet and introducing light travel time delays as well as an obscuring torus is, on its own, not enough to induce a global bipolar asymmetry. We carried out a complementary study of the allowed parameter space and found consistent parameters for both jets, supporting our conclusion of symmetry. However, while this is the result of doing the analysis over an extended period of time, asymmetries may still arise at individual time steps. Studies focusing on inducing asymmetries in the jet launching stage could give us further insight into the morphology and evolution of jets, and the astrophysical processes behind them. 
Important new constraints will come from future observational facilities at high radio frequencies such as the next-generation Event Horizon Telescope \citep[ngEHT;][]{Doeleman2023}
and the next-generation Very Large Array \citep[ngVLA;][]{murphy18,Kadler2023}.

In the presented work, we used 2D hydrodynamic simulations. Besides the suppression of 3D instabilities due to the assumed axisymmetry of the jets, there are other limitations to our model, such as the lack of magnetic fields in the hydrodynamic simulations. The effects of magnetic fields can be split into two categories: the impact on the dynamics and the impact on the computed emission. If we assume that the jets are not magnetically dominated on parsec scales, the sound speed will be larger than the Alfv\'{e}n speed. In this case, the oscillation of the jet surface will be determined by the sound speed and we expect no significant alteration of the jet shape compared to pure SRHD simulations. On the other hand, if the jets are magnetically dominated the Alfv\'{e}n speed could be larger than the sound speed. In this case, the wavelength of the oscillations will decrease and a more cylindrical collimation profile ($k=0$) could be obtained \citep[see, e.g.,][]{Porth2015}. Given the cylindrical collimation profile found in some observations, this scenario could be analyzed in future studies but is out of the scope of this work. 
On the emission side, magnetic fields can introduce a top-bottom asymmetry in relativistic jets due to the aberration of light and pitch angle of the magnetic field \citep[see, e.g.,][]{Fuentes2018}. While this aberration is typically small for large inclinations and small velocities, and has not been studied in this work, the presence of magnetic fields could introduce a general asymmetry in a double-sided jet system.

\begin{acknowledgements}
%\section{Acknowledgments}
It is a pleasure to thank L. Rezzolla for useful comments and suggestions. This research is supported by the DFG research grant ``Jet physics on horizon scales and beyond" (Grant No.  443220636) within the DFG research unit ``Relativistic Jets in Active Galaxies" (FOR 5195). Support also comes from the European Research Council Advanced Grant “JETSET: Launching, propagation and emission of relativistic jets from binary mergers and across mass scales” (grant no. 884631). The numerical simulations and calculations have been performed on \texttt{MISTRAL} at the Chair of Astronomy at the JMU Wuerzburg and on \texttt{Iboga} and \texttt{Calea} at the Institute for Theoretical Physics (ITP) at University of Frankfurt. ASP and MP acknowledge support by the Spanish Ministry of Science through Grant and PID2022-136828NB-C43. MP acknowledges support from the Generalitat Valenciana through grant CIPROM/2022/49, and from the Astrophysics and High Energy Physics programme supported by the Ministry of Science and Innovation and Generalitat Valenciana with funding from European Union NextGenerationEU (PRTR-C17.I1) through grant ASFAE/2022/005. YM is supported by the National Key R\&D Program of China (grant no. 2023YFE0101200), the National Natural Science Foundation of China (grant no. 12273022), and the Shanghai municipality orientation program of basic research for international scientists (grant no. 22JC1410600).
\end{acknowledgements}

\bibliographystyle{aa} 
\bibliography{bibliography}

\begin{thebibliography}{55}
\expandafter\ifx\csname natexlab\endcsname\relax\def\natexlab#1{#1}\fi

\bibitem[{{Agudo} {et~al.}(2001){Agudo}, {G{\'o}mez}, {Mart{\'\i}},
  {Ib{\'a}{\~n}ez}, {Marscher}, {Alberdi}, {Aloy}, \& {Hardee}}]{Agudo2001}
{Agudo}, I., {G{\'o}mez}, J.-L., {Mart{\'\i}}, J.-M., {et~al.} 2001, \apjl,
  549, L183

\bibitem[{Allan {et~al.}(2024)Allan, Caswell, Keim, van~der Wel, \&
  Verweij}]{allan_2024_10696534}
Allan, D.~B., Caswell, T., Keim, N.~C., van~der Wel, C.~M., \& Verweij, R.~W.
  2024, soft-matter/trackpy: v0.6.2

\bibitem[{{Aloy} {et~al.}(2003){Aloy}, {Mart{\'\i}}, {G{\'o}mez}, {Agudo},
  {M{\"u}ller}, \& {Ib{\'a}{\~n}ez}}]{Aloy2003}
{Aloy}, M.-{\'A}., {Mart{\'\i}}, J.-M., {G{\'o}mez}, J.-L., {et~al.} 2003,
  \apjl, 585, L109

\bibitem[{{Baczko}(2020)}]{2020PhDT........17B}
{Baczko}, A.-K. 2020, PhD thesis, Max-Planck-Institute for Radioastronomy, Bonn

\bibitem[{{Baczko} {et~al.}(2022){Baczko}, {Ros}, {Kadler}, {Fromm},
  {Boccardi}, {Perucho}, {Krichbaum}, {Burd}, \&
  {Zensus}}]{2022A&A...658A.119B}
{Baczko}, A.~K., {Ros}, E., {Kadler}, M., {et~al.} 2022, \aap, 658, A119

\bibitem[{{Baczko} {et~al.}(2019){Baczko}, {Schulz}, {Kadler}, {Ros},
  {Perucho}, {Fromm}, \& {Wilms}}]{2019A&A...623A..27B}
{Baczko}, A.~K., {Schulz}, R., {Kadler}, M., {et~al.} 2019, \aap, 623, A27

\bibitem[{{Baczko} {et~al.}(2016){Baczko}, {Schulz}, {Kadler}, {Ros},
  {Perucho}, {Krichbaum}, {B{\"o}ck}, {Bremer}, {Grossberger}, {Lindqvist},
  {Lobanov}, {Mannheim}, {Mart{\'\i}-Vidal}, {M{\"u}ller}, {Wilms}, \&
  {Zensus}}]{2016A&A...593A..47B}
{Baczko}, A.~K., {Schulz}, R., {Kadler}, M., {et~al.} 2016, \aap, 593, A47

\bibitem[{{Blandford} {et~al.}(2019){Blandford}, {Meier}, \&
  {Readhead}}]{2019ARA&A..57..467B}
{Blandford}, R., {Meier}, D., \& {Readhead}, A. 2019, \araa, 57, 467

\bibitem[{{Blandford} \& {Payne}(1982)}]{1982MNRAS.199..883B}
{Blandford}, R.~D. \& {Payne}, D.~G. 1982, \mnras, 199, 883

\bibitem[{{Blandford} \& {Znajek}(1977)}]{1977MNRAS.179..433B}
{Blandford}, R.~D. \& {Znajek}, R.~L. 1977, \mnras, 179, 433

\bibitem[{{Boccardi} {et~al.}(2021){Boccardi}, {Perucho}, {Casadio}, {Grandi},
  {Macconi}, {Torresi}, {Pellegrini}, {Krichbaum}, {Kadler}, {Giovannini},
  {Karamanavis}, {Ricci}, {Madika}, {Bach}, {Ros}, {Giroletti}, \&
  {Zensus}}]{2021A&A...647A..67B}
{Boccardi}, B., {Perucho}, M., {Casadio}, C., {et~al.} 2021, \aap, 647, A67

\bibitem[{{B{\"o}ck}(2013)}]{2013PhDT.......479B}
{B{\"o}ck}, M. 2013, PhD thesis, University of Erlangen-Nuremberg, Astronomical
  Institute

\bibitem[{{Chael} {et~al.}(2023){Chael}, {Issaoun}, {Pesce}, {Johnson},
  {Ricarte}, {Fromm}, \& {Mizuno}}]{2023ApJ...945...40C}
{Chael}, A., {Issaoun}, S., {Pesce}, D.~W., {et~al.} 2023, \apj, 945, 40

\bibitem[{{Chael} {et~al.}(2018){Chael}, {Johnson}, {Bouman}, {Blackburn},
  {Akiyama}, \& {Narayan}}]{2018ApJ...857...23C}
{Chael}, A.~A., {Johnson}, M.~D., {Bouman}, K.~L., {et~al.} 2018, \apj, 857, 23

\bibitem[{{Chael} {et~al.}(2016){Chael}, {Johnson}, {Narayan}, {Doeleman},
  {Wardle}, \& {Bouman}}]{2016ApJ...829...11C}
{Chael}, A.~A., {Johnson}, M.~D., {Narayan}, R., {et~al.} 2016, \apj, 829, 11

\bibitem[{{Chashkina} {et~al.}(2021){Chashkina}, {Bromberg}, \&
  {Levinson}}]{2021MNRAS.508.1241C}
{Chashkina}, A., {Bromberg}, O., \& {Levinson}, A. 2021, \mnras, 508, 1241

\bibitem[{{Doeleman} {et~al.}(2023){Doeleman}, {Barrett}, {Blackburn},
  {Bouman}, {Broderick}, {Chaves}, {Fish}, {Fitzpatrick}, {Freeman}, {Fuentes},
  {G{\'o}mez}, {Haworth}, {Houston}, {Issaoun}, {Johnson}, {Kettenis},
  {Loinard}, {Nagar}, {Narayanan}, {Oppenheimer}, {Palumbo}, {Patel}, {Pesce},
  {Raymond}, {Roelofs}, {Srinivasan}, {Tiede}, {Weintroub}, \&
  {Wielgus}}]{Doeleman2023}
{Doeleman}, S.~S., {Barrett}, J., {Blackburn}, L., {et~al.} 2023, Galaxies, 11,
  107

\bibitem[{{Fendt} \& {Sheikhnezami}(2013)}]{2013ApJ...774...12F}
{Fendt}, C. \& {Sheikhnezami}, S. 2013, \apj, 774, 12

\bibitem[{{Fichet de Clairfontaine} {et~al.}(2022){Fichet de Clairfontaine},
  {Meliani}, \& {Zech}}]{2022A&A...661A..54F}
{Fichet de Clairfontaine}, G., {Meliani}, Z., \& {Zech}, A. 2022, \aap, 661,
  A54

\bibitem[{{Fosbury} {et~al.}(1978){Fosbury}, {Mebold}, {Goss}, \&
  {Dopita}}]{1978MNRAS.183..549F}
{Fosbury}, R.~A.~E., {Mebold}, U., {Goss}, W.~M., \& {Dopita}, M.~A. 1978,
  \mnras, 183, 549

\bibitem[{{Fromm} {et~al.}(2016){Fromm}, {Perucho}, {Mimica}, \&
  {Ros}}]{2016A&A...588A.101F}
{Fromm}, C.~M., {Perucho}, M., {Mimica}, P., \& {Ros}, E. 2016, \aap, 588, A101

\bibitem[{{Fromm} {et~al.}(2018){Fromm}, {Perucho}, {Porth}, {Younsi}, {Ros},
  {Mizuno}, {Zensus}, \& {Rezzolla}}]{2018A&A...609A..80F}
{Fromm}, C.~M., {Perucho}, M., {Porth}, O., {et~al.} 2018, \aap, 609, A80

\bibitem[{{Fromm} {et~al.}(2019){Fromm}, {Younsi}, {Baczko}, {Mizuno}, {Porth},
  {Perucho}, {Olivares}, {Nathanail}, {Angelakis}, {Ros}, {Zensus}, \&
  {Rezzolla}}]{2019A&A...629A...4F}
{Fromm}, C.~M., {Younsi}, Z., {Baczko}, A., {et~al.} 2019, \aap, 629, A4

\bibitem[{{Fuentes} {et~al.}(2018){Fuentes}, {G{\'o}mez}, {Mart{\'\i}}, \&
  {Perucho}}]{Fuentes2018}
{Fuentes}, A., {G{\'o}mez}, J.~L., {Mart{\'\i}}, J.~M., \& {Perucho}, M. 2018,
  \apj, 860, 121

\bibitem[{{G{\'o}mez} {et~al.}(1997){G{\'o}mez}, {Mart{\'\i}}, {Marscher},
  {Ib{\'a}{\~n}ez}, \& {Alberdi}}]{1997ApJ...482L..33G}
{G{\'o}mez}, J.~L., {Mart{\'\i}}, J.~M., {Marscher}, A.~P., {Ib{\'a}{\~n}ez},
  J.~M., \& {Alberdi}, A. 1997, \apjl, 482, L33

\bibitem[{{Gomez} {et~al.}(1995){Gomez}, {Marti}, {Marscher}, {Ibanez}, \&
  {Marcaide}}]{1995ApJ...449L..19G}
{Gomez}, J.~L., {Marti}, J.~M.~A., {Marscher}, A.~P., {Ibanez}, J.~M.~A., \&
  {Marcaide}, J.~M. 1995, \apjl, 449, L19

\bibitem[{{Hada} {et~al.}(2013){Hada}, {Kino}, {Doi}, {Nagai}, {Honma},
  {Hagiwara}, {Giroletti}, {Giovannini}, \& {Kawaguchi}}]{2013ApJ...775...70H}
{Hada}, K., {Kino}, M., {Doi}, A., {et~al.} 2013, \apj, 775, 70

\bibitem[{{Ho} {et~al.}(1997){Ho}, {Filippenko}, \&
  {Sargent}}]{1997ApJS..112..315H}
{Ho}, L.~C., {Filippenko}, A.~V., \& {Sargent}, W. L.~W. 1997, \apjs, 112, 315

\bibitem[{{Jensen} {et~al.}(2003){Jensen}, {Tonry}, {Barris}, {Thompson},
  {Liu}, {Rieke}, {Ajhar}, \& {Blakeslee}}]{2003ApJ...583..712J}
{Jensen}, J.~B., {Tonry}, J.~L., {Barris}, B.~J., {et~al.} 2003, \apj, 583, 712

\bibitem[{{Jiang} {et~al.}(2023){Jiang}, {Mizuno}, {Fromm}, \&
  {Nathanail}}]{Jiang2023}
{Jiang}, H.-X., {Mizuno}, Y., {Fromm}, C.~M., \& {Nathanail}, A. 2023, \mnras,
  522, 2307

\bibitem[{{Kadler} {et~al.}(2004{\natexlab{a}}){Kadler}, {Kerp}, {Ros},
  {Falcke}, {Pogge}, \& {Zensus}}]{2004A&A...420..467K}
{Kadler}, M., {Kerp}, J., {Ros}, E., {et~al.} 2004{\natexlab{a}}, \aap, 420,
  467

\bibitem[{{Kadler} {et~al.}(2023){Kadler}, {Riechers}, {Agarwal}, {Baczko},
  {Beuther}, {Bigiel}, {Birnstiel}, {Boccardi}, {Bomans}, {Boogaard}, {Braun},
  {Britzen}, {Br{\"u}ggen}, {Brunthaler}, {Caselli}, {Els{\"a}sser}, {von
  Fellenberg}, {Flock}, {Fromm}, {Fuhrmann}, {Hartogh}, {Hoeft}, {Keenan},
  {Kovalev}, {Kreckel}, {Livingston}, {Lobanov}, {M{\"u}ller}, {Ros},
  {Schilke}, {De Simone}, {Spitler}, {Ueda}, {Vardoulaki}, {Vegetti}, {Weis},
  {Wendel}, {Xu}, {Zhao}, {Albrecht}, {Basu}, {Becker Tjus}, {Bernhart},
  {Blum}, {Bonnassieux}, {Bredendiek}, {van Delden}, {Di Gennaro}, {Enders},
  {Eppel}, {Hase}, {Hoang}, {Hugentobler}, {Kaasinen}, {Krupp}, {Kun},
  {Laubach}, {Lin}, {Mannheim}, {Menten}, {Perkuhn}, {Pohl}, {Powell},
  {Rezzolla}, {Ricci}, {Schinnerer}, {Schmidt}, {Sch{\"o}pfel}, {Stanko},
  {Stein}, {Sulzenauer}, {Taziaux}, {Tursunov}, {Walter}, {Weiss}, {Witzel},
  {Wolf}, {Zensus}, {Mus}, {Toth}, {Alberdi}, {Benisty}, {Cox}, {Guirado},
  {Johnson}, {Juvela}, {Neeleman}, {Pashchenko}, {P{\'e}rez Torres}, {Perraut},
  \& {Zajacek}}]{Kadler2023}
{Kadler}, M., {Riechers}, D.~A., {Agarwal}, J., {et~al.} 2023, arXiv e-prints,
  arXiv:2311.10056

\bibitem[{{Kadler} {et~al.}(2004{\natexlab{b}}){Kadler}, {Ros}, {Lobanov},
  {Falcke}, \& {Zensus}}]{2004A&A...426..481K}
{Kadler}, M., {Ros}, E., {Lobanov}, A.~P., {Falcke}, H., \& {Zensus}, J.~A.
  2004{\natexlab{b}}, \aap, 426, 481

\bibitem[{{Kameno} {et~al.}(2003){Kameno}, {Inoue}, {Wajima}, {Sawada-Satoh},
  \& {Shen}}]{2003PASA...20..134K}
{Kameno}, S., {Inoue}, M., {Wajima}, K., {Sawada-Satoh}, S., \& {Shen}, Z.-Q.
  2003, \pasa, 20, 134

\bibitem[{{Kameno} {et~al.}(2020){Kameno}, {Sawada-Satoh}, {Impellizzeri},
  {Espada}, {Nakai}, {Sugai}, {Terashima}, {Kohno}, {Lee}, \&
  {Mart{\'\i}n}}]{2020ApJ...895...73K}
{Kameno}, S., {Sawada-Satoh}, S., {Impellizzeri}, C.~M.~V., {et~al.} 2020,
  \apj, 895, 73

\bibitem[{{Komissarov} \& {Falle}(1996)}]{1996ASPC..100..173K}
{Komissarov}, S.~S. \& {Falle}, S.~A.~E.~G. 1996, in Astronomical Society of
  the Pacific Conference Series, Vol. 100, Energy Transport in Radio Galaxies
  and Quasars, ed. P.~E. {Hardee}, A.~H. {Bridle}, \& J.~A. {Zensus}, 173

\bibitem[{{Komissarov} \& {Falle}(1997)}]{1997MNRAS.288..833K}
{Komissarov}, S.~S. \& {Falle}, S.~A.~E.~G. 1997, \mnras, 288, 833

\bibitem[{{Kovalev} {et~al.}(2020){Kovalev}, {Pushkarev}, {Nokhrina}, {Plavin},
  {Beskin}, {Chernoglazov}, {Lister}, \& {Savolainen}}]{2020MNRAS.495.3576K}
{Kovalev}, Y.~Y., {Pushkarev}, A.~B., {Nokhrina}, E.~E., {et~al.} 2020, \mnras,
  495, 3576

\bibitem[{{Lindeberg}(1994)}]{1994JApSt..21..225L}
{Lindeberg}, T. 1994, Journal of Applied Statistics, 21, 225

\bibitem[{{Lister} {et~al.}(2009){Lister}, {Aller}, {Aller}, {Cohen}, {Homan},
  {Kadler}, {Kellermann}, {Kovalev}, {Ros}, {Savolainen}, {Zensus}, \&
  {Vermeulen}}]{2009AJ....137.3718L}
{Lister}, M.~L., {Aller}, H.~D., {Aller}, M.~F., {et~al.} 2009, \aj, 137, 3718

\bibitem[{{Mahlmann} {et~al.}(2020){Mahlmann}, {Levinson}, \&
  {Aloy}}]{2020MNRAS.494.4203M}
{Mahlmann}, J.~F., {Levinson}, A., \& {Aloy}, M.~A. 2020, \mnras, 494, 4203

\bibitem[{{Mayall}(1939)}]{1939PASP...51..282M}
{Mayall}, N.~U. 1939, \pasp, 51, 282

\bibitem[{{Mimica} {et~al.}(2009){Mimica}, {Aloy}, {Agudo}, {Mart{\'\i}},
  {G{\'o}mez}, \& {Miralles}}]{2009ApJ...696.1142M}
{Mimica}, P., {Aloy}, M.~A., {Agudo}, I., {et~al.} 2009, \apj, 696, 1142

\bibitem[{{Murphy} {et~al.}(2018){Murphy}, {Bolatto}, {Chatterjee}, {Casey},
  {Chomiuk}, {Dale}, {de Pater}, {Dickinson}, {Francesco}, {Hallinan},
  {Isella}, {Kohno}, {Kulkarni}, {Lang}, {Lazio}, {Leroy}, {Loinard},
  {Maccarone}, {Matthews}, {Osten}, {Reid}, {Riechers}, {Sakai}, {Walter}, \&
  {Wilner}}]{murphy18}
{Murphy}, E.~J., {Bolatto}, A., {Chatterjee}, S., {et~al.} 2018, in
  Astronomical Society of the Pacific Conference Series, Vol. 517, Science with
  a Next Generation Very Large Array, ed. E.~{Murphy}, 3

\bibitem[{{Nakahara} {et~al.}(2020){Nakahara}, {Doi}, {Murata}, {Nakamura},
  {Hada}, {Asada}, {Sawada-Satoh}, \& {Kameno}}]{2020AJ....159...14N}
{Nakahara}, S., {Doi}, A., {Murata}, Y., {et~al.} 2020, \aj, 159, 14

\bibitem[{{Nathanail} {et~al.}(2020){Nathanail}, {Fromm}, {Porth}, {Olivares},
  {Younsi}, {Mizuno}, \& {Rezzolla}}]{2020MNRAS.495.1549N}
{Nathanail}, A., {Fromm}, C.~M., {Porth}, O., {et~al.} 2020, \mnras, 495, 1549

\bibitem[{{Parfrey} {et~al.}(2015){Parfrey}, {Giannios}, \&
  {Beloborodov}}]{2015MNRAS.446L..61P}
{Parfrey}, K., {Giannios}, D., \& {Beloborodov}, A.~M. 2015, \mnras, 446, L61

\bibitem[{{Perucho} {et~al.}(2010){Perucho}, {Mart{\'\i}}, {Cela}, {Hanasz},
  {de La Cruz}, \& {Rubio}}]{Perucho2010}
{Perucho}, M., {Mart{\'\i}}, J.~M., {Cela}, J.~M., {et~al.} 2010, \aap, 519,
  A41

\bibitem[{{Porth} \& {Komissarov}(2015)}]{Porth2015}
{Porth}, O. \& {Komissarov}, S.~S. 2015, \mnras, 452, 1089

\bibitem[{{Ricci} {et~al.}(2022){Ricci}, {Boccardi}, {Nokhrina}, {Perucho},
  {MacDonald}, {Mattia}, {Grandi}, {Madika}, {Krichbaum}, \&
  {Zensus}}]{2022A&A...664A.166R}
{Ricci}, L., {Boccardi}, B., {Nokhrina}, E., {et~al.} 2022, \aap, 664, A166

\bibitem[{{Thompson} {et~al.}(2017){Thompson}, {Moran}, \&
  {Swenson}}]{2017isra.book.....T}
{Thompson}, A.~R., {Moran}, J.~M., \& {Swenson}, George~W., J. 2017,
  {Interferometry and Synthesis in Radio Astronomy, 3rd Edition}

\bibitem[{van~der Walt {et~al.}(2014)van~der Walt, Schönberger,
  Nunez-Iglesias, Boulogne, Warner, Yager, Gouillart, Yu, \& the scikit-image
  contributors}]{scikit-image}
van~der Walt, S., Schönberger, J.~L., Nunez-Iglesias, J., {et~al.} 2014,
  PeerJ, 2, e453

\bibitem[{{Vermeulen} {et~al.}(2003){Vermeulen}, {Ros}, {Kellermann}, {Cohen},
  {Zensus}, \& {van Langevelde}}]{2003A&A...401..113V}
{Vermeulen}, R.~C., {Ros}, E., {Kellermann}, K.~I., {et~al.} 2003, \aap, 401,
  113

\bibitem[{{Woo} \& {Urry}(2002)}]{2002ApJ...579..530W}
{Woo}, J.-H. \& {Urry}, C.~M. 2002, \apj, 579, 530

\bibitem[{{Wrobel}(1984)}]{1984ApJ...284..531W}
{Wrobel}, J.~M. 1984, \apj, 284, 531

\end{thebibliography}

\begin{appendix}

\section{Tracking results}

\begin{figure}[h]
\centering
\includegraphics[width=9cm]{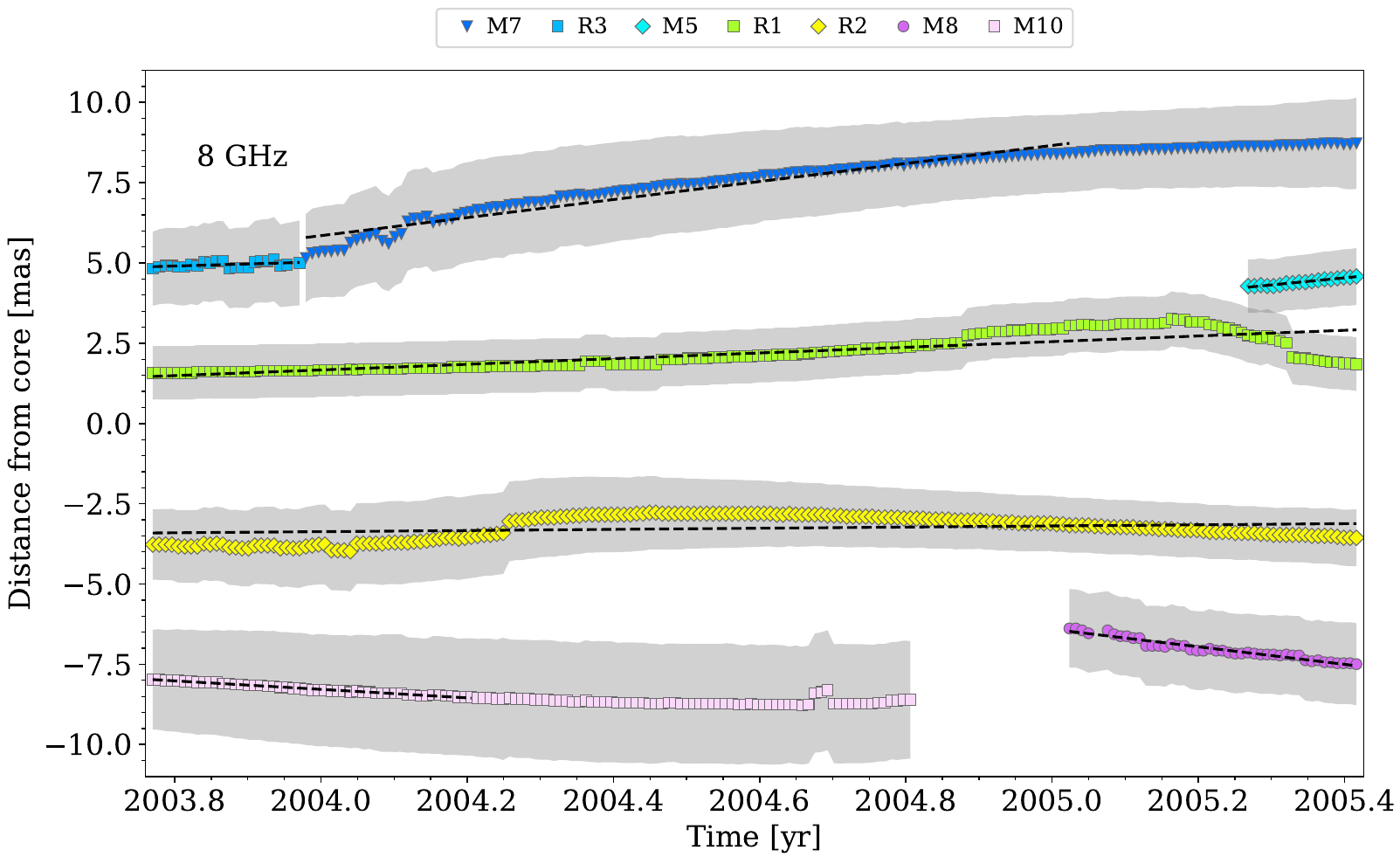}
\caption{Distance of bright components from the central core over time for the 8~GHz maps. The components are marked by a letter and a number as explained in Fig.~\ref{fig:component43}. The dashed lines on each component correspond to a linear fit.}
\label{fig:component8}
\end{figure}

\begin{figure}[h]
\centering
\includegraphics[width=9cm]{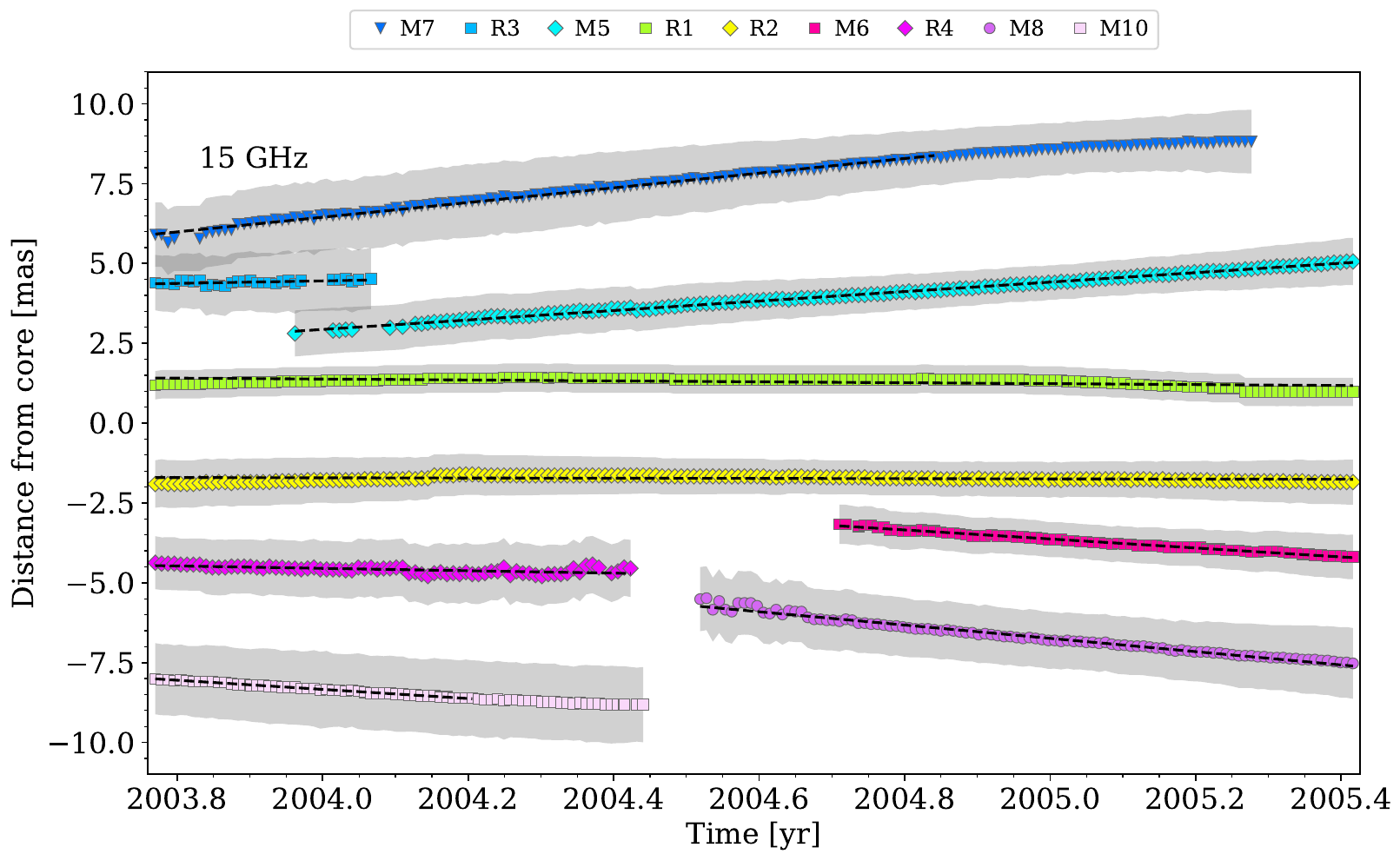}
\caption{Distance of bright components from the central core over time for the 15~GHz maps. The dashed lines on each component correspond to a linear fit.}
\label{fig:component15}
\end{figure}

\begin{figure}[h]
\centering
\includegraphics[width=9cm]{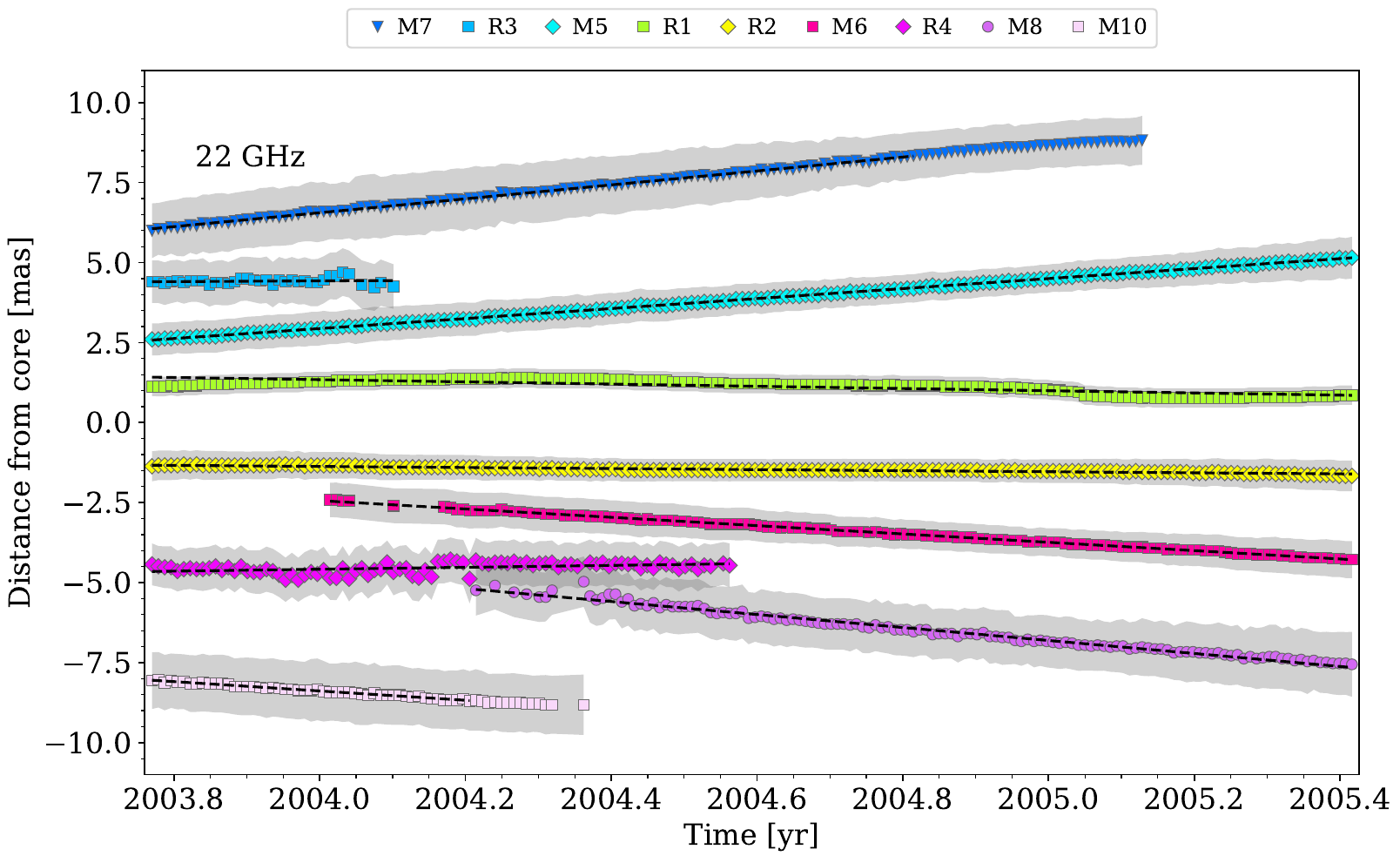}
\caption{Distance of bright components from the central core over time for the 22~GHz maps. The dashed lines on each component correspond to a linear fit.}
\label{fig:component22}
\end{figure}

Figs.~\ref{fig:component8}, \ref{fig:component15} and \ref{fig:component22} show the component trajectories for 8~GHz, 15~GHz and 22~GHz, respectively. These have been computed with the same method described in Section~\ref{kinematics}.

\begin{figure}[h!]
\centering
\includegraphics[width=9cm]{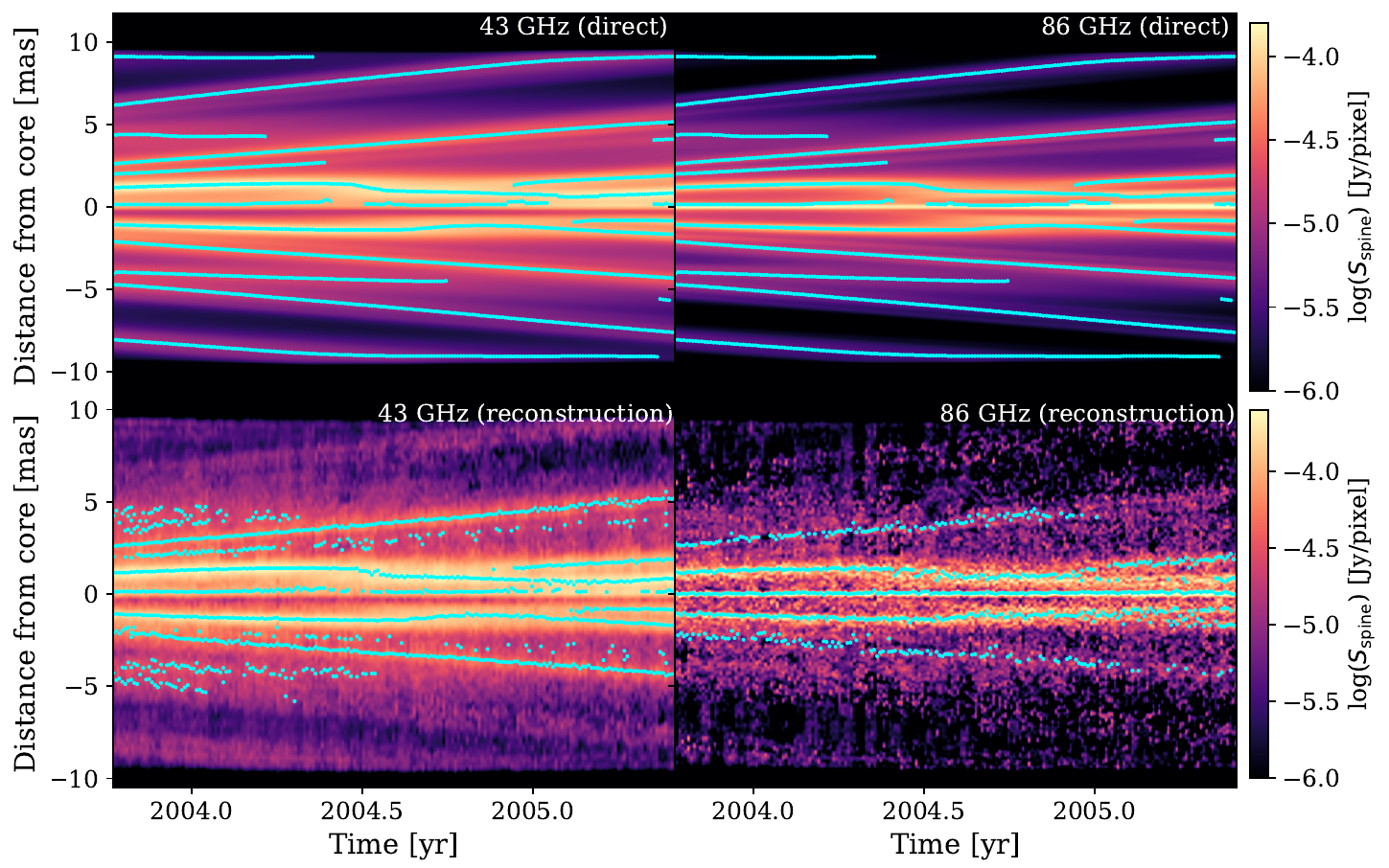}
\caption{Evolution over time of the flux of the jet spine. Each vertical slice shows the flux of the jet spine for a given time step, following the positional angle of 15$^{\circ}$, showing the evolution of the jets' shock structure over time. The direct emission maps (top) have been convolved with the same beam as the corresponding reconstructed maps (bottom). Overplotted in cyan are the components detected by separately applying our tracking algorithm to each of the maps. Note the effect of adjacent shocks in the tracking algorithm, such as the oscillation of the recollimation shocks when a moving shock passes through, or the jet core temporarily going undetected in the direct maps when the recollimation shock at $\sim$1~mas is dragged outward and expands. Note also the faint trailing components visible in the direct images. The direct emission maps also show the purely numerical effect of the moving shocks ``slowing down" as they reach the end of the simulated box.}
\label{fig:timespaceplot}
\end{figure}

Fig.~\ref{fig:timespaceplot} compares the results of applying our tracking algorithm on the reconstructed images (bottom) and the direct, blurred ones (top) for 43~GHz and 86~GHz. The minimum detection threshold was increased for the reconstructed images due to the noise present at these frequencies.

\section{Fitting results} \label{appendixfitting}

Here we provide a more careful study of the results of fitting Eq.~\ref{eq:broken_power_law} to our width profiles.

\begin{figure}[h]
\centering
\includegraphics[width=9cm]{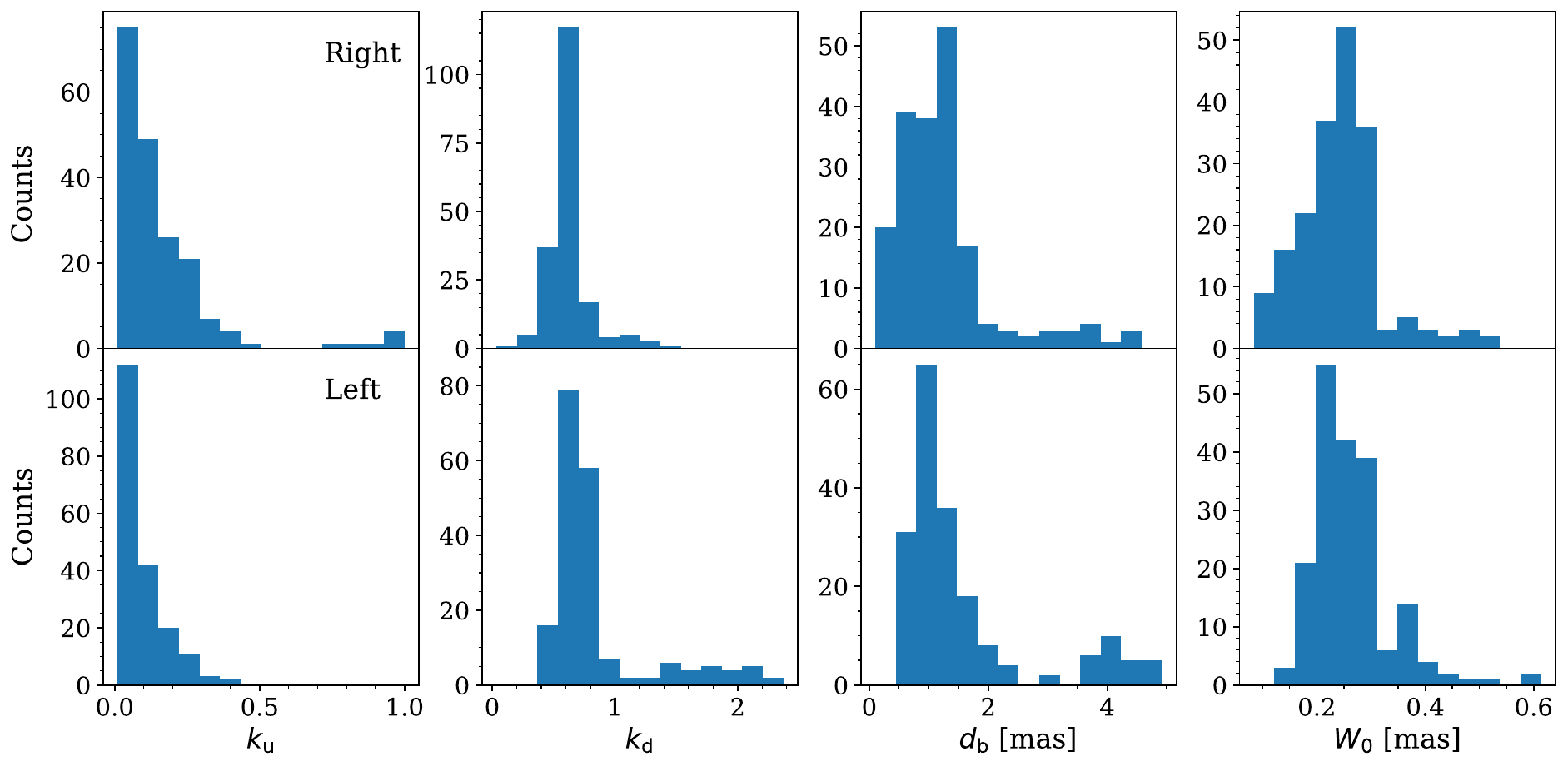}
\caption{Distribution of parameters resulting from fitting a broken power-law to our reconstructed maps. Note the tail present on the distribution of the breakpoints, $d_{\textrm{b}}$.}
\label{fig:fitting_hist}
\end{figure}

Fig.~\ref{fig:fitting_hist} shows the resulting distribution of all four fitting parameters, for the left and right jets. The downstream slopes, $k_{\textrm{d}}$, are the same ones found in Fig.~\ref{fig:kde}. Before the breakpoint, the fitting results in a less than parabolic profile ($k_{\textrm{u}}<0.5$) in $>90\%$ of the time steps, for either jet or counterjet, while $k_{\textrm{u}}<0.25$ for $>80\%$ of time steps. After the breakpoint, a downstream value of $k_{\textrm{d}}>1$ was found in 5\% of the time steps for the right jet and in 15\% for the left jet. This results in a tail following the central distribution.

\begin{figure}[h]
\centering
\includegraphics[width=8cm]{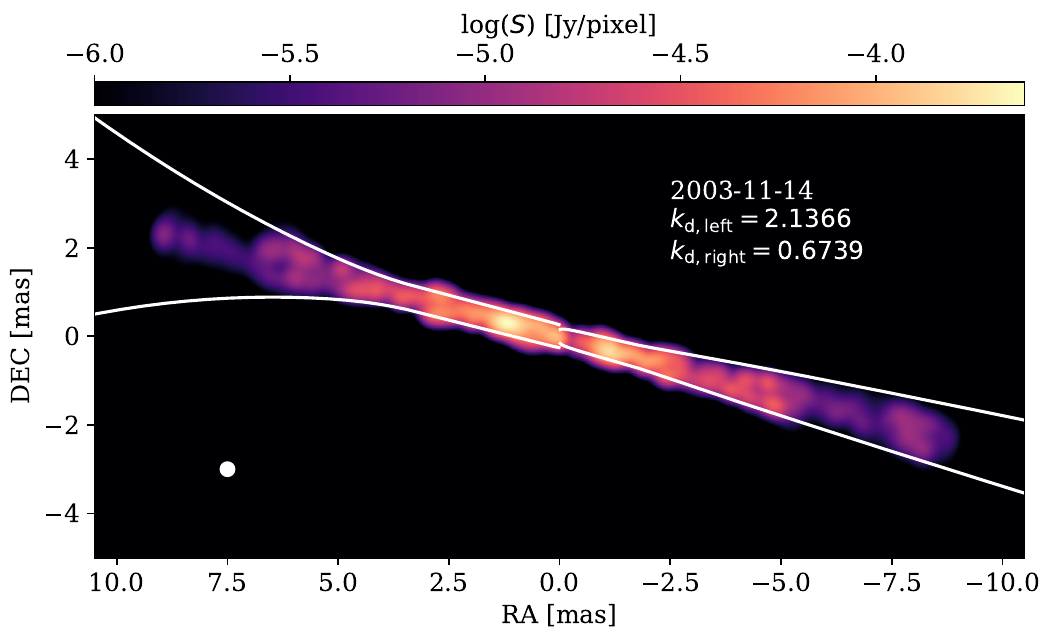}
\caption{Emission map at 43\,GHz showing the overshot fitting of the downstream eastern jet. The solid white line represents the broken power-law fit of this specific time step.}
\label{fig:highfit}
\end{figure}

These greater than conical values could have resulted from the breakpoint being placed further out than the primary recollimation shocks during the fitting process. We checked this by plotting the result of our fitting on top of its corresponding emission map at 43~GHz. Figure~\ref{fig:highfit} shows an example of such a reconstruction. On the left jet, the breakpoint was placed at a distance of $d_{\textrm{b,l}}=3.94$~mas from the core. It appears likely that missing information on particularly faint outer features leads to the breakpoint being identified with a cylindrical region at this distance. When extended farther out, we clearly observe the failed fitting overestimates the width at the final few mas of the jet, caused by a greater than conical expansion ($k_{\textrm{d}}>1$). For these reasons, we decided to re-run the analysis in Section~\ref{collimation} removing all values $k_{\textrm{d}}>1.5$.

\begin{figure}[h]
\centering
\includegraphics[width=9cm]{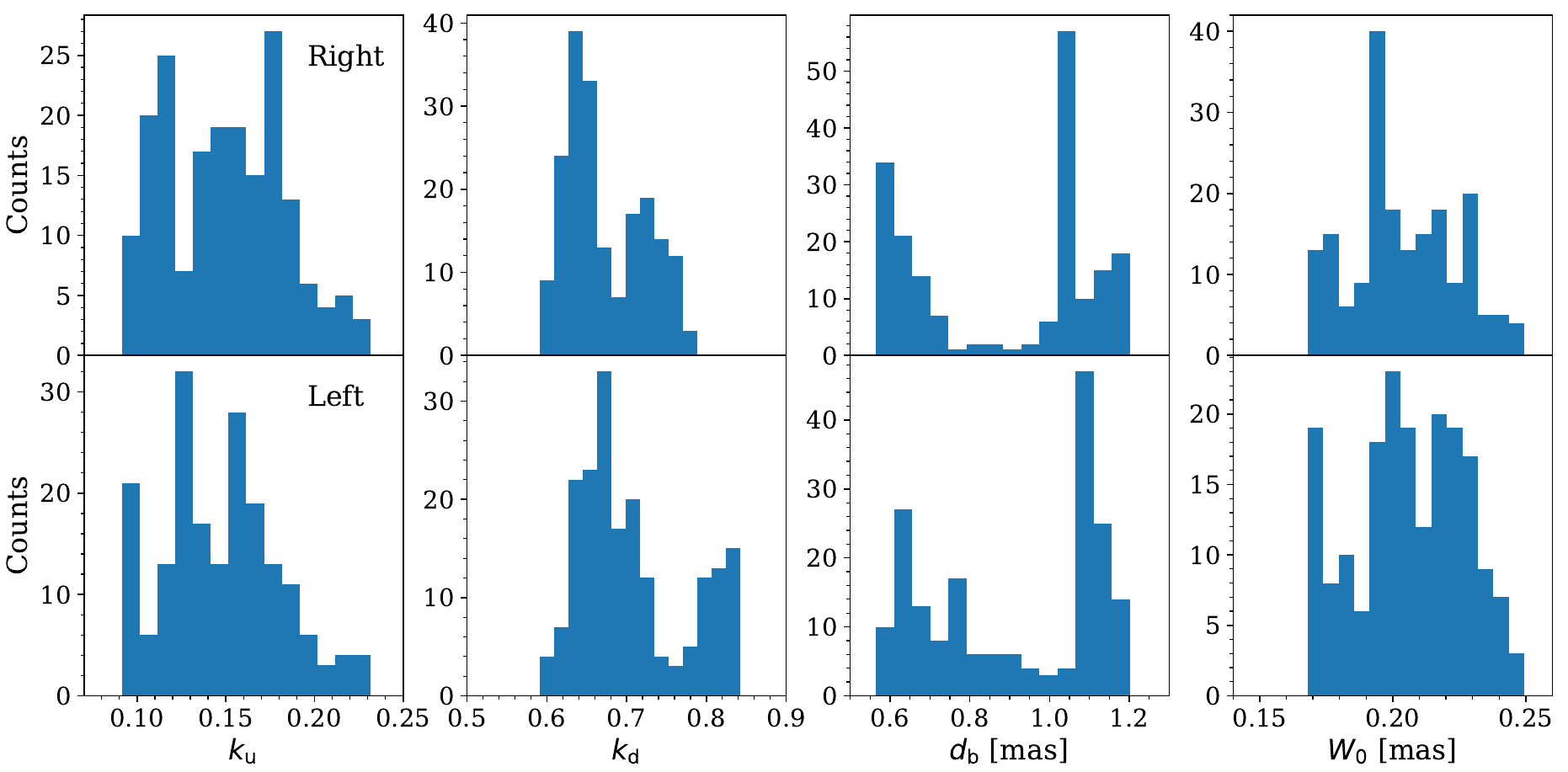}
\caption{Distribution of parameters resulting from fitting a broken power-law to our direct, infinite resolution maps. Note the narrower range of values when compared to Fig.~\ref{fig:fitting_hist}.}
\label{fig:theory_fitting_hist}
\end{figure}

\begin{figure}[h]
\centering
\includegraphics[width=9cm]{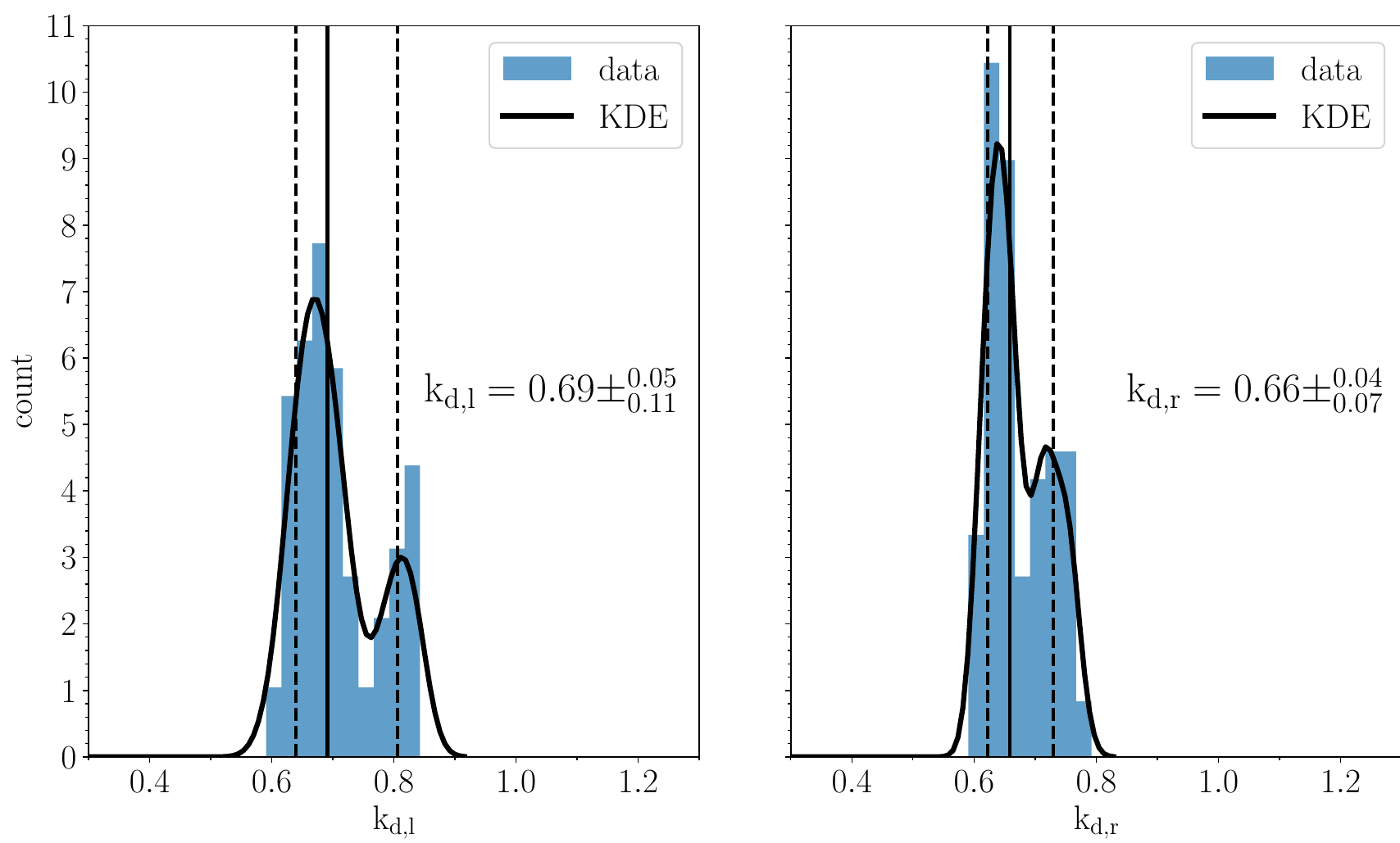}
\caption{Kernel density estimation of the distribution of downstream slopes resulting from the broken power-law fitting.}
\label{fig:kde_theory}
\end{figure}

To further validate our analysis, we ran our fitting algorithm on the direct emission maps. The fitting parameters are shown in Fig.~\ref{fig:theory_fitting_hist}. Note the difference in x-axis values when compared to Fig.~\ref{fig:fitting_hist}. The resulting fitting values were constrained to a smaller range for every parameter. The breakpoints do not show a tail in their distribution, with their maximum distance found to be $d_{\textrm{b,l}}=1.2$~mas. This is accompanied by every value of the downstream slope being less than conical, $k_{\textrm{d}}<1$. In Fig.~\ref{fig:kde_theory}, we ran the same analysis shown in Fig.~\ref{fig:kde} with this data. The resulting median values support our conclusion that there is no global asymmetry present, in agreement with our intrinsically symmetric simulation. This analysis also shows the impact which a finite-resolution array has on the study of jet kinematics.

\begin{figure}[h]
	\centering
	\includegraphics[width=8cm]{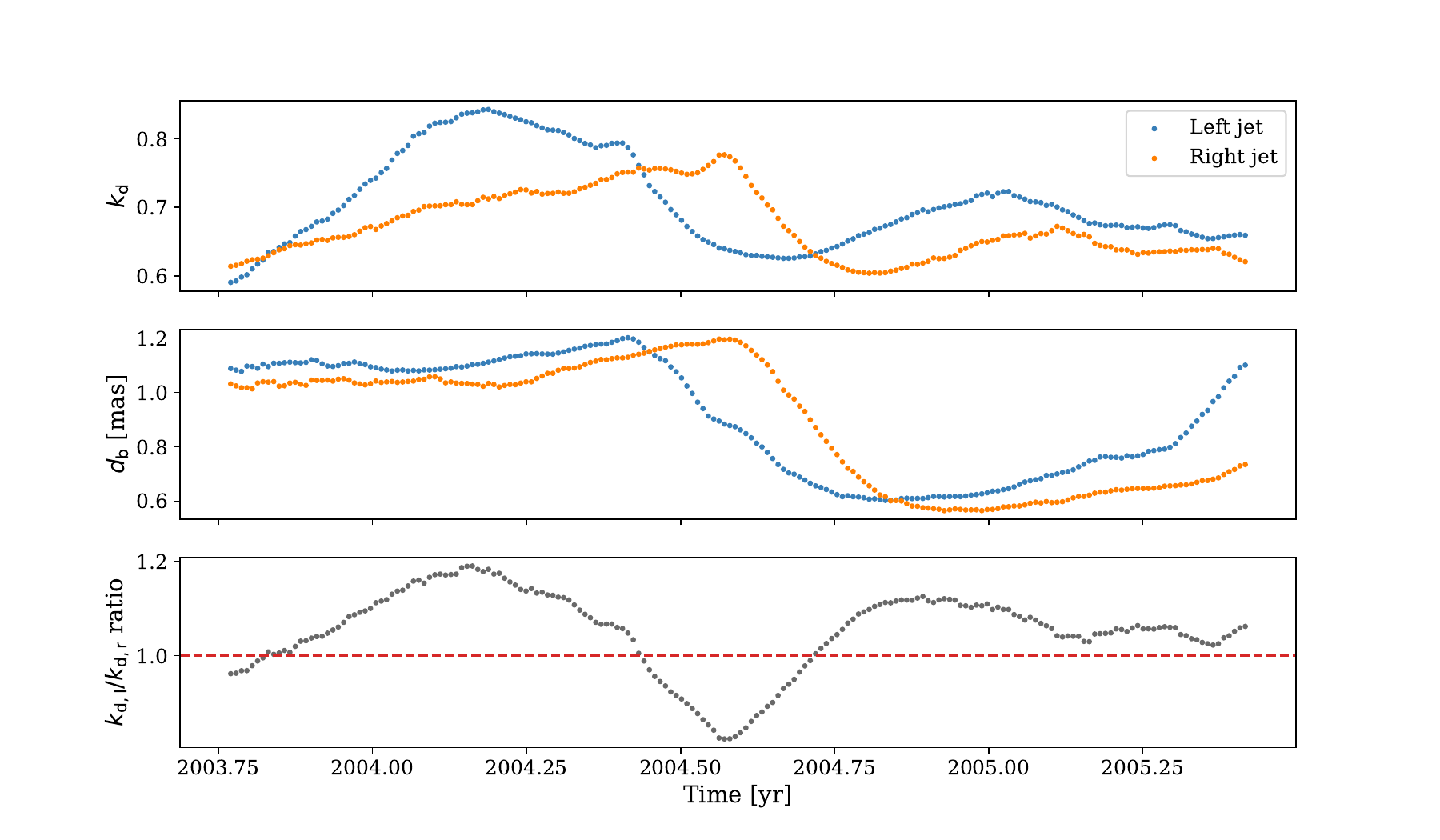}
	\caption{Parameters resulting from fitting a broken power-law to our infinite resolution maps, as a function of time. The downstream slopes for both jets (top panel) and the ratio between them (bottom panel) are contrasted to the breakpoint location (central panel). A dashed line in the bottom panel indicates a ratio of 1.}
	\label{fig:ratios}
\end{figure}

Finally, Fig.~\ref{fig:ratios} shows some of the parameters of the infinite-resolution fitting of Fig.~\ref{fig:theory_fitting_hist} as a function of time. We focus on the infinite-resolution maps as the reconstruction process introduces significant scatter and makes it challenging to find such clear trends, though the overall distributions point to similar results as seen in Figs.~\ref{fig:kde} and \ref{fig:kde_theory}. The results of Fig.~\ref{fig:ratios} show the dependence between the collimation profile and the breakpoint location, where an inwards displacement of $d_{\textrm{b}}$ seems to be accompanied by a narrower collimation profile. Furthermore, when contrasting this to Fig.~\ref{fig:timespaceplot}, the dates where both values of $d_{\textrm{b}}$ drop roughly correspond to the dates where the recollimation shocks at $\pm 1$~mas display inward motion. As previously discussed, this oscillation is caused by the recollimation shock being dragged downstream by a moving shock. This emphazises the importance of travelling components on the overall jet profile, as they may introduce asymmetries on specific epochs when combined with observational effects.
\end{appendix}
\end{document}